\documentclass[twocolumn, journal]{IEEEtran}
\IEEEoverridecommandlockouts

\usepackage{color}
\usepackage{graphicx}
\usepackage{epstopdf}
\usepackage{amsmath}
\usepackage{amssymb}
\usepackage{algorithm}
\usepackage{algorithmic}
\usepackage{amsmath}
\usepackage{multirow}
\usepackage{booktabs}
\usepackage{array}
\usepackage{amsthm}
\usepackage{stfloats}
\usepackage{caption}
\usepackage{subfigure}
\usepackage{bm}
\usepackage{setspace}
\usepackage{diagbox}
\usepackage{enumitem}

\newcommand{\be}{\begin{equation}}
\newcommand{\ee}{\end{equation}}
\newcommand{\bea}{\begin{eqnarray}}
\newcommand{\eea}{\end{eqnarray}}
\newcommand{\ba}{\begin{array}}
\newcommand{\ea}{\end{array}}

\allowdisplaybreaks[4]
\usepackage{flushend}

\newcommand{\tabincell}[2]{\begin{tabular}{@{}#1@{}}#2\end{tabular}}
\def\BibTeX{{\rm B\kern-.05em{\sc i\kern-.025em b}\kern-.08em
    T\kern-.1667em\lower.7ex\hbox{E}\kern-.125emX}}
\begin{document}

\title{Dynamic Hybrid Beamforming Designs for ELAA Near-Field Communications
\thanks{Part of this paper has been presented at the IEEE International Conference on Communications (ICC), Denver, CO, USA, June 2024 \cite{1}.
 }
\thanks{M. Liu and M. Li are with the School of Information and Communication Engineering, Dalian University of Technology, Dalian 116024, China (e-mail: liumengzhen@mail.dlut.edu.cn, mli@dlut.edu.cn).}
\thanks{R. Liu was with the School of Information and Communication Engineering, Dalian University of Technology, Dalian 116024, China. She is currently with the Center for Pervasive Communications and Computing, University of California, Irvine, CA 92697, USA (e-mail: rangl2@uci.edu).}
\thanks{Q. Liu is with the School of Computer Science and Technology, Dalian University of Technology, Dalian 116024, China (e-mail: qianliu@dlut.edu.cn).}
}

\author{Mengzhen Liu,~\IEEEmembership{Student Member,~IEEE,}
        Ming Li,~\IEEEmembership{Senior Member,~IEEE,}
        Rang Liu,~\IEEEmembership{Member,~IEEE,}\\
        and Qian Liu,~\IEEEmembership{Member,~IEEE}\vspace{-0.5cm}}
\maketitle
\pagestyle{empty}
\thispagestyle{empty}

\begin{abstract}
Extremely large-scale antenna array (ELAA) is a key candidate technology for the sixth generation (6G) mobile networks. Nevertheless, using substantial numbers of antennas to transmit high-frequency signals in ELAA systems significantly exacerbates the near-field effect. Unfortunately, traditional hybrid beamforming schemes are highly vulnerable to ELAA near-field communications. To effectively mitigate severe near-field effect, we propose a novel dynamic hybrid beamforming architecture for ELAA systems, in which each antenna is either adaptively connected to one radio frequency (RF) chain for signal transmission or deactivated for power saving. For the case that instantaneous channel state information (CSI) is available during each channel coherence time, a real-time dynamic hybrid beamforming design is developed to maximize the achievable sum rate under the constraints of the constant modulus of phase-shifters (PSs), non-overlapping dynamic connection network and total transmit power. When instantaneous CSI cannot be easily obtained in real-time, we propose a two-timescale dynamic hybrid beamforming design, which optimizes analog beamformer in long-timescale and digital beamformer in short-timescale, with the goal of maximizing ergodic sum-rate under the same constraints. Simulation results demonstrate the advantages of the proposed dynamic hybrid beamforming architecture and the effectiveness of the developed algorithms for ELAA near-field communications.
\end{abstract}


\section{Introduction}
The forthcoming sixth generation (6G) wireless communication system is expected to utilize ultra-wide spectrum resources within the millimeter wave (mmWave) and terahertz (THz) frequency bands to facilitate data transmission rates of up to 1 Tbps \cite{W. Saad 2020}. Despite its beneficial features, the signal attenuation becomes increasingly severe as signal frequency rises, posing a challenge to signal propagation.
The emerging extremely large-scale antenna array (ELAA) technology presents a promising solution to address the challenge of severe path loss of mmWave/THz signals  \cite{E. Bjornson 2019}.
As a natural evolution of massive multiple-input multiple-output (MIMO), ELAA augments the number of antennas by at least an order of magnitude thanks to the rapid development of hardware technology. In addition to compensating for path loss, ELAA provides unprecedented high spectral efficiency, large coverage area, strong interference mitigation, and fine spatial resolution for 6G wireless communications. These advantages fulfill a number of key performance indicator requirements, including peak data rate, reliability, and positioning accuracy \cite{C. You 2023}-\cite{M. Cui 2022}.
It is foreseeable that ELAA will become a pivotal enabling technology in shaping the landscape of future 6G networks.

While enjoying its benefits, ELAA operating in high-frequency bands will unavoidably confront two issues in practical applications: The requirement for hybrid beamforming architecture in hardware implementation and the emergence of near-field effect in channel characteristics. Firstly, the increased number of antennas and higher operating frequencies will incur exorbitant hardware expenses and power consumption if employing a conventional fully-digital beamforming architecture \cite{A. F. Molisch 2017}.
To tackle the hardware limitations while maintaining satisfactory performance, the current literature generally suggests implementing the ELAA with the hybrid beamforming architecture, which employs only a few expensive radio frequency (RF) chains for digital beamformer to enable multiplexing transmissions and utilizes a large number of hardware-efficient phase-shifters (PSs) for analog beamformer to compensate for the severe path loss \cite{C. Han 2021}. To strike a balance between the performance and hardware complexity, two classic hybrid architectures have been extensively investigated for mmWave/THz massive MIMO systems: Fully-connected hybrid architecture where each RF chain connects to all antennas \cite{F. Sohrabi 2015}, \cite{Z. Wang 2018} and fixed-subarray hybrid architecture where each RF chain only connects to a fixed subset of antennas \cite{L. Dai 2016}.

Secondly, using more antennas for transmitting higher frequency signals will extend the Fresnel (near-field) region and consequently present a new challenge, i.e., exacerbating the near-field effect \cite{{L. Dai 2022 NF}}, \cite{C. You 2023 Mixed}.
This phenomenon can be observed in three distinct aspects: \textit{i}) The relationship between the phase and antenna index is nonlinear; \textit{ii}) the path loss differs among antennas due to variations in propagation distance; and \textit{iii}) the radiated power varies across antennas due to differences in angle of departure (AoD).
Specifically, researchers have recently explored near-field  characteristics and revealed that the phase of the antenna index transitions from linear to nonlinear. Instead of utilizing the far-field uniform plane wave (UPW) channel model, the near-field uniform spherical wave (USW) channel model is introduced in \cite{M. Cui 2022}. With the further increase of the number of antennas, the non-uniform spherical wave (NUSW) model is presented in \cite{Y. Zeng 2022} to appropriately describe the signal amplitude variations caused by different path losses across the array elements. Based on this NUSW model, performance analysis for ELAA systems is conducted in \cite{Y. Zeng 2021}, \cite{E. Bjornson 2020}. Instead of considering different distances between each antenna and the user, the radiation pattern of the antenna and different AoDs are considered in \cite{T. Cui 2021}.

As ELAA generally comes with hybrid beamforming and near-field communications, it is worth noting that the traditional hybrid beamforming architecture is vulnerable to near-field effect, which may lead to performance degradation \cite{Y. Zeng 2023}-\cite{Y. Zeng SNR}.
Specifically, conventional fully-connected and fixed-subarray hybrid beamforming structures can only adjust the phases without changing the amplitudes of signals transmitted from all antennas, and thus cannot adapt to the near-field channel where the channel gains for each antenna are different.
Moreover, as the number of antennas increases, the difference between the distributions of transmitting signal power and channel gain across the entire array will become more pronounced, inevitably reducing beamforming gain and wasting more power.
Thus, sophisticated hybrid beamforming architectures should be explored for ELAA near-field communications to overcome the near-field effect.

The dynamic hybrid beamforming architecture has recently emerged as a strategy to offer more flexibility by adaptively adjusting the mapping between antennas and RF chains and/or selectively activating RF chains \cite{Y. Cai antenna selection}-\cite{F. Yang 2020}. Several studies have already verified its potential to provide performance gains in ELAA near-field communication systems. To be specific, the authors in \cite{Y. Liu 2023 dynamic}, \cite{L. Dai 2022} introduce a relatively simple architecture that can only configure each RF chain to be either active or inactive, aiming at achieving the spatial multiplexing-power consumption tradeoff when accounting for the nonlinear phase of near-field effect. However, the simple dynamic architectures in \cite{Y. Liu 2023 dynamic} and \cite{L. Dai 2022} limits the design flexibility and hinders the potential performance improvement.
Motivated by these findings, we aim to fully exploit the flexibility of dynamic hybrid beamforming architecture to compensate for the performance loss caused by the near-field effect.

It also should be noted that all the aforementioned dynamic hybrid beamforming designs rely on the knowledge of the instantaneous channel state information (CSI) when optimizing both analog and digital beamformers. However, within ELAA systems, the substantial number of antennas and the hybrid beamforming architectures will provoke significant training/signaling complexity. Thus, the significant overhead of channel estimation renders the real-time-CSI-based hybrid beamforming design impractical in a rapidly changing wireless environment. Recently, a two-timescale hybrid beamforming design scheme has been proposed to overcome this issue, where the analog beamformer is designed based on channel statistics in the long-timescale, while the digital beamformer is designed with instantaneous low-dimensional effective CSI in the short-timescale \cite{A. Liu two-stage}. This framework is attractive in some practical scenarios since it significantly reduces the computational complexity and system overhead, and substantially improves system efficiency, thereby having been extensively investigated for conventional hybrid beamforming designs in \cite{Y. Cai 2022}-\cite{Y. Cai 2020}. Unfortunately, its application in dynamic hybrid beamforming design for near-field communications remains blank.

Motivated by the above discussions, in this paper, we investigate the dynamic hybrid beamforming designs under real-time and two-timescale frameworks for ELAA near-field communications. The main contributions of this paper can be summarized as follows:
\begin{itemize}
  \item We first comprehensively analyze the channel characteristics of ELAA systems, specifically focusing on the nonlinear phase, as well as the varied channel gains due to different distances and AoDs across antennas, which will inevitably cause severe near-field effect. In order to mitigate it, we introduce a novel dynamic hybrid beamforming architecture, where each antenna is allowed to either dynamically connect to one RF chain for transmitting signal, or not connect to any RF chain (i.e., switched to ``off mode") for power saving, which offers more design flexibility for performance improvement.
  \item When the instantaneous CSI can be effectively acquired during each channel coherence time, a real-time dynamic hybrid beamforming algorithm is developed. Specifically, we aim to jointly design the analog and digital beamformers to maximize the achievable sum-rate under the constraints of the constant modulus of PSs, non-overlapping dynamic connection network, and the total transmit power. To tackle the resulting complicated non-convex problem, we employ the fractional programming (FP) method to transform the objective function into a more tractable form and then exploit the manifold-based algorithm to tackle the non-convex constraint.
  \item When the instantaneous CSI cannot be easily obtained in real-time, a two-timescale dynamic hybrid beamforming algorithm is proposed to reduce channel estimation overhead. The analog beamformer is optimized in the long-timescale, and the digital beamformer is designed in the short-timescale, with the purpose of maximizing the ergodic sum rate under the same constraints. Stochastic successive convex approximation (SSCA) is employed to substitute the stochastic non-convex objective with a convex quadratic surrogate function. The majorization-minimization (MM) method, along with logarithmic surrogate functions, is applied to convert the binary constraints into continuous, smooth, and convex constraints.
  \item Extensive simulation results are provided to verify the capability of the proposed dynamic hybrid beamforming architecture to alleviate the near-field effect in the ELAA systems as well as the effectiveness of the proposed two dynamic hybrid beamforming design algorithms.
\end{itemize}

\textit{Notations}: Boldface lower-case and upper-case letters indicate column vectors and matrices, respectively. $(\cdot)^{*}$, $(\cdot)^{T}$, $(\cdot)^{H}$, and $(\cdot)^{-1}$ denote the conjugate, transpose, transpose-conjugate, and inverse operations, respectively. $\mathrm{vec}(\mathbf{A})$ vectorizes the matrix $\mathbf{A}$. $\angle a$ is the angle of complex-valued $a$. $|a|$, $\|\mathbf{a}\|$, $\|\mathbf{a}\|_{0}$, and $\|\mathbf{A}\|_{F}$ are the magnitude of scalar $a$, the norm of vector $\mathbf{a}$, the $l_{0}$-norm of vector $\mathbf{a}$, and the Frobenius norm of matrix $\mathbf{A}$. $\mathrm{diag}\{\mathbf{a}\}$ indicates the diagonal matrix whose diagonals are the elements of $\mathbf{a}$. $\mathbf{I}_{M}$ represents a $M\times M$ identity matrix. $\Re\{\cdot\}$ denotes the real part of a complex number. $\mathbb{C}$ and $\mathbb{R}$ denote the sets of complex numbers and real numbers, respectively. $\mathbb{E}\{\cdot\}$ represents statistical expectation. Notations $\odot$ and $\otimes$ are the Hadamard product and Kronecker product of matrices, respectively. Finally, $\mathbf{A}(i,:)$ and $\mathbf{A}(i,j)$ denote the $i$-th row and $(i,j)$-th element of matrix $\mathbf{A}$, respectively.

\vspace{-0.3cm}
\section{System Model}
\begin{figure}[!t]
  \centering
  \includegraphics[width= 3.4 in]{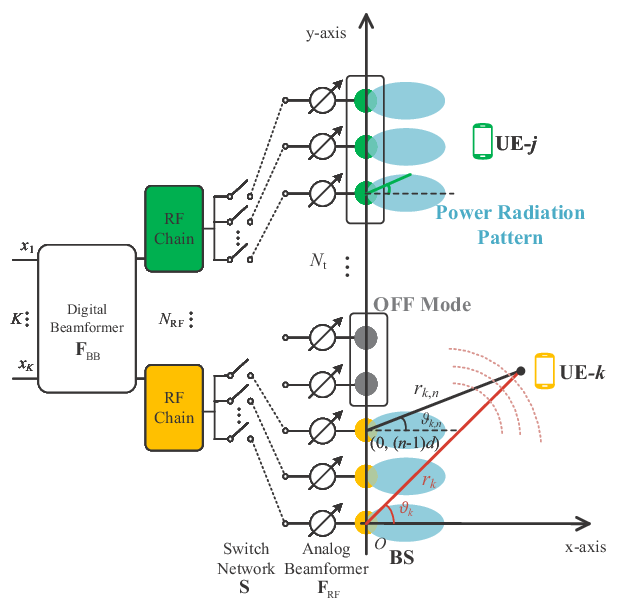}
  \caption{ELAA near-field communication with dynamic hybrid beamforming.}
  \label{fig:systemmodel}
\end{figure}

We consider a mmWave downlink communication system as illustrated in Fig. \ref{fig:systemmodel}, where a base station (BS) equipped with an extremely large-scale uniform linear array (ULA) serves multiple single-antenna user equipments (UEs). The antenna spacing is $d={\lambda_{\mathrm{c}}}/{2}$, where $\lambda_{\mathrm{c}}$ is the center carrier wavelength.
Let $\mathcal{N}\triangleq \{1, 2, \ldots, N_{\mathrm{t}}\}$ denote the set of BS antennas, and $\mathcal{K}\triangleq \{1, 2, \ldots, K\}$ denote the set of UEs.
It is assumed that the UEs are located in the near-field region of the ELAA, as shown in Fig. \ref{fig:systemmodel}. By placing the origin of the coordinate system at the bottom of the ELAA as a reference point, the coordinate of the $n$-th antenna can be written as $(0,(n-1)d)$. The $k$-th UE is located at $(r_{k}\cos\vartheta_{k},r_{k}\sin\vartheta_{k})$, where $r_{k}$ and $\vartheta_{k}$ denote the distance and the AoD from the reference point of ELAA to the $k$-th UE, respectively. Furthermore, the distance between the $n$-th antenna and the $k$-th UE can be derived as
\begin{equation}
\begin{split}
    r_{k,n}&=\sqrt{(r_{k}\cos\vartheta_{k})^{2}+(r_{k}\sin\vartheta_{k}-(n-1)d)^{2}}\\
    &=\sqrt{r_{k}^{2}+((n\!-\!1)d)^{2}-2(n\!-\!1)dr_{k}\sin\vartheta_{k}},~\forall k, n.
\end{split}
\end{equation}
The AoD from the $n$-th antenna to the $k$-th UE can be calculated as
\begin{equation}
\vartheta_{k,n}=\arccos\left(\frac{r_{k}\cos\vartheta_{k}}{r_{k,n}}\right),~\forall k, n.
\end{equation}

As previously emphasized, the accurate modeling of near-field channels in the ELAA system requires careful consideration of three essential factors outlined as follows:
\begin{itemize}
\item The phase of the spherical wave is a nonlinear function of the antenna index, which is determined by both distance and angle.
\item The path loss is distinct from each antenna element to the UE due to distance variations, which can be expressed as $L_{k,n}=10^{-\frac{C_{0}}{10}}(\frac{r_{k,n}}{D_{0}})^{-\alpha}$, where $C_{0}$ denotes the signal attenuation at the reference distance $D_{0}$, and $\alpha$ denotes the path loss exponent.
\item The radiation gain is varied across the entire antenna array due to different angles between each antenna element and the UE, which can be written as $G_{k,n}=(\cos\vartheta_{k,n})^{3},~\vartheta_{k,n}\in (-\frac{\pi}{2},\frac{\pi}{2})$.
\end{itemize}

After carefully examining the aforementioned factors, the near-field channel between the ELAA and the $k$-th UE can be characterized in a more accurate model as
\begin{equation}
\mathbf{h}_{k}\!\!\triangleq\!\!\big[g_{k,1}e^{-\jmath\frac{2\pi}{\lambda_{\mathrm{c}}}r_{k,1}}, g_{k,2}e^{-\jmath\frac{2\pi}{\lambda_{\mathrm{c}}}r_{k,2}}\!,\cdots\!, g_{k,N_{\mathrm{t}}}\!e^{-\jmath\frac{2\pi}{\lambda_{\mathrm{c}}}r_{k,N_{\mathrm{t}}}}\!\big]^{T}\!\!,~\forall k,
\end{equation}
where $g_{k,n} = \sqrt{L_{k,n}G_{k,n}}$ is the channel gain between the $n$-th antenna and the $k$-th UE.

After presenting a comprehensive near-field channel model incorporating the three factors mentioned above, our focus now shifts toward investigating the beamforming design for this ELAA near-field communication system.
The BS is equipped with an ELAA and adopts a hybrid beamforming architecture to reduce hardware complexity and power consumption.
Conventional hybrid beamforming architectures (e.g., full-connected and fixed-subarray) only allow signal phase-shift adjustment and have uniform power distribution across antennas.
As a result, ELAA systems using such conventional hybrid beamforming will be highly vulnerable to the near-field effect.
To be more specific, the antenna element positioned at a relatively larger distance and/or larger AoD from the UE results in weaker receiving signals due to its poorer near-field channel gain.
Moreover, it is important to note that this antenna element still consumes the same amount of transmit power as the other elements despite the weaker reception at the UE.
The existence of these ``ineffective'' antennas leads to wasted power, which may reduce beamforming gain as the number of antennas increases in near-field communication systems.

In an effort to mitigate the performance degradation due to the near-field effect, we propose a novel dynamic hybrid beamforming architecture, as shown in Fig. \ref{fig:systemmodel}.
Specifically, each antenna is either activated by dynamically connecting to one RF chain for signal transmission, or deactivated as not being driven by any RF chain for power saving.
In other words, each RF chain is adaptively connected to a disjoint set of antennas via a switch network for dynamically forming a non-overlapping subarray, meanwhile some ``ineffective" antennas are switched to the ``off mode''.
By fully utilizing the design flexibility offered by dynamic antenna selection and activation, the proposed dynamic hybrid beamforming architecture has the potential to significantly enhance the performance of ELAA near-field communications.

Without loss of generality, in this paper, we assume that the number of RF chains $N_{\mathrm{RF}}$ is greater than or equal to the number of UEs $K$, i.e., $N_{\mathrm{RF}}\geq K$, and let $\mathcal{N}_{\mathrm{RF}}\triangleq \{1, 2, \ldots, N_{\mathrm{RF}}\}$ denote the set of RF chains. In the proposed dynamic hybrid architecture, the analog beamformer is defined as $\mathbf{F}_{\mathrm{RF}}\in \{\mathcal{F},0\}^{N_{\mathrm{t}}\times N_{\mathrm{RF}}}$, where $\mathcal{F}\triangleq \{\phi=\frac{1}{\sqrt{N_{\mathrm{t}}}}e^{\jmath\theta}|\theta\in [0, 2\pi)\}$ is the continuous phase set of the PSs following the constant modulus constraint. Such a definition implies that if the $l$-th RF chain is connected to the $n$-th antenna, the phase-shift corresponding to the $n$-th antenna has a non-zero value, i.e., $\mathbf{F}_{\mathrm{RF}}(n,l)\in \mathcal{F}$; if not, it is set as zero, i.e., $\mathbf{F}_{\mathrm{RF}}(n,l)=0$. Moreover, each antenna will be dynamically connected to only one RF chain or remain ``off mode". This implies that the analog beamformer matrix will have either one or zero non-zero elements in each row, i.e., $\|\mathbf{F}_{\mathrm{RF}}(n,:)\|_{0} \in \{1, 0\}$.

Based on the above mathematical definition, the signal received by the $k$-th UE is given by
\begin{equation}
y_{k} =\mathbf{h}_{k}^{H}\mathbf{F}_{\mathrm{RF}}\mathbf{F}_{\mathrm{BB}}\mathbf{x}+n_{k},~ \forall k,
\end{equation}
where $\mathbf{h}_{k}\!\in \!\mathbb{C}^{N_{\mathrm{t}}\times 1}$ denotes the downlink channel between the BS and the $k$-th UE, $\mathbf{F}_{\mathrm{BB}}\triangleq[\mathbf{f}_{\mathrm{BB},1}, \mathbf{f}_{\mathrm{BB},2}, \ldots, \mathbf{f}_{\mathrm{BB},K}]\in \mathbb{C}^{N_{\mathrm{RF}}\times K}$ denotes the digital beamformer, $\mathbf{x}\triangleq[x_{1}, x_2, \ldots, x_{K}]\in \mathbb{C}^{K\times 1}$ denotes the transmitted symbols of $K$ UEs which satisfies $\mathbb{E}\{\mathbf{x}\mathbf{x}^{H}\}=\mathbf{I}_{K}$, and $n_{k}\!\sim\!\mathcal{CN}(0,\sigma^{2}_{k})$ denotes the additive white Gaussian noise (AWGN) at the $k$-th UE.
The signal-to-interference-plus-noise ratio (SINR) of the $k$-th UE can be calculated as
\begin{equation}
\mathrm{SINR}_{k}=\frac{|\mathbf{h}_{k}^{H}\mathbf{F}_{\mathrm{RF}}\mathbf{f}_{\mathrm{BB},k}|^{2}}{\sum_{j=1,j\neq k}^{K}|\mathbf{h}_{k}^{H}\mathbf{F}_{\mathrm{RF}}\mathbf{f}_{\mathrm{BB},j}|^{2}+\sigma_{k}^{2}}, ~\forall k.\\
\end{equation}
And the corresponding achievable rate can be derived as $R_{k}=\log_{2}(1+\mathrm{SINR}_{k})$.
In the following sections, we will tackle the near-field effect through dynamic hybrid beamforming designs, based on real-time and two-timescale frameworks, respectively. In particular, when the instantaneous CSI is easily accessible, the real-time framework can be employed; conversely, if the instantaneous CSI is challenging to acquire, the two-timescale framework offers a more suitable solution.

\section{Real-time Dynamic Hybrid Beamforming Design}
\subsection{Problem Formulation}
In this section, we assume that the BS has perfect instantaneous CSI of all near-field UEs during each channel coherence time. The instantaneous CSI can be obtained by utilizing a time-division duplex (TDD) operation assuming the uplink-downlink channel reciprocity according to the existing near-field channel estimation and beam training methods \cite{M. Cui 2022}, \cite{C. You Fast near-field beam training}, \cite{C. You beam management}. Based on the acquired instantaneous CSI, we aim to jointly design the analog beamformer $\mathbf{F}_{\mathrm{RF}}$ and the digital beamformer $\mathbf{F}_{\mathrm{BB}}$ to maximize the real-time achievable sum-rate under the constraints of the constant modulus of PSs, non-overlapping dynamic connection network, and the total transmit power. Therefore, the real-time dynamic hybrid beamforming design problem can be formulated as
\begin{subequations}
\label{eq:obj}
\begin{align}
    \label{eq:obj_a}
    \max_{\mathbf{F}_{\mathrm{RF}}, \mathbf{F}_{\mathrm{BB}}}&\sum^{K}_{k=1}R_{k}\\
    \label{eq:obj_b}
    \mathrm{s.t.}\ \ &\mathbf{F}_{\mathrm{RF}}(n,l)\in \{\mathcal{F},0\}, ~\forall n,l,\\
    \label{eq:obj_c}
    &\|\mathbf{F}_{\mathrm{RF}}(n,:)\|_{0}\in\{1,0\}, ~\forall n,\\
    \label{eq:obj_d} &\|\mathbf{F}_{\mathrm{RF}}\mathbf{F}_{\mathrm{BB}}\|_{F}^{2}= P_{\mathrm{t}},
\end{align}
\end{subequations}
where $P_{\mathrm{t}}$ is the total transmit power.

The optimization problem \eqref{eq:obj} is challenging due to the fractional and $\log(\cdot)$ terms in the objective function \eqref{eq:obj_a}, the non-convex constraints of the analog beamformer, and the power constraint. Therefore, in this section, we first employ the FP method to convert the complicated objective function \eqref{eq:obj_a} into a solvable polynomial expression. Then, the dynamic analog beamformer matrix $\mathbf{F}_{\mathrm{RF}}$ is decomposed into a diagonal phase-shift matrix $\mathbf{A}$ and a switch matrix $\mathbf{S}$. Based on such decomposition, we convert the design of a dynamic analog beamformer into two more tractable subproblems, one of which involves solving $\mathbf{A}$ by manifold algorithm. Finally, efficient algorithms are developed to solve two subproblems and design the digital beamformer $\mathbf{F}_{\mathrm{BB}}$ iteratively.

\vspace{-0.3cm}
\subsection{FP-based Transformation of Objective Function}
We begin with utilizing the Lagrangian dual transformation to convert the original objective function \eqref{eq:obj_a} into a more tractable form. To be specific, by substituting each ratio term $\mathrm{SINR}_{k}$ in the objective function \eqref{eq:obj_a} with the auxiliary variables $\{\mu_{k}\}_{k=1}^{K}$, the unconstrained sum-rate maximization problem can be equivalently reformulated as
\begin{subequations}
\begin{align}
    \max_{\{\mu_{k}\}_{k=1}^{K}}&\ \ \sum_{k=1}^{K}\log(1+\mu_{k})\\
    \mathrm{s.t. }\ \ &\ \ \ \mu_{k}\leq \mathrm{SINR}_{k}, ~\forall k,
\end{align}
\end{subequations}
which is a convex problem. By integrating the inequality constraints into the objective function, we can obtain the Lagrangian function as
\begin{equation}
\label{Lagrangian function}
 \mathcal{L} = \sum_{k=1}^{K}\log(1+\mu_{k})- \sum_{k=1}^{K}\lambda_{k}(\mu_{k}-\mathrm{SINR}_{k}),
\end{equation}
where $\{\lambda_{k}\}_{k=1}^{K}$ are multipliers for the constraints. Then, the optimal solutions of $\{\mu_{k},\lambda_{k}\}_{k=1}^{K}$  can be determined by setting the partial derivative of $\mathcal{L}$ with respect to them to zero, which yield
\begin{equation}
\label{muSINR}
 \mu_{k}^{\star} = \mathrm{SINR}_{k},~\forall k,
\end{equation}
\begin{equation}
\label{lambda}
 \lambda_{k}^{\star} = \frac{1}{1+\mu_{k}^{\star}},~\forall k.
\end{equation}
With the optimal value of $\{\mu_{k}^{\star},\lambda_{k}^{\star}\}_{k=1}^{K}$ given in \eqref{muSINR}, \eqref{lambda}, the maximum of the Lagrangian function in \eqref{Lagrangian function} can be calculated as
\begin{equation}
\begin{aligned}
 \mathcal{L}^\star &=\sum_{k=1}^K \log(1 + \mu_{k}^{\star}) - \sum_{k=1}^K \frac{1}{1+\mu_{k}^{\star}}(\mu_{k}^{\star} - \mathrm{SINR}_k)\\
&= \sum_{k = 1}^K \log(1 + \mu_{k}^{\star}) - \sum_{k = 1}^K \mu_{k}^{\star} \\
 &~~~~+ \sum_{k = 1}^K \frac{(1 + \mu_{k}^{\star})|\mathbf{h}_k^H \mathbf{F}_{\mathrm{RF}}\mathbf{f}_{\mathrm{BB},k}|^2}{\sum_{j = 1}^K |\mathbf{h}_k^H \mathbf{F}_{\mathrm{RF}}\mathbf{f}_{\mathrm{BB},j}|^2 + \sigma_k^2}.\\
\end{aligned}
\end{equation}
Therefore, the transformed objective function can be reformulated as
\begin{equation} \label{eq:sum log}
\sum_{k=1}^{K}\log_{2}(1\!+\!\mu_{k})-\!\!\sum_{k=1}^{K}\mu_{k}+\sum_{k=1}^{K}\frac{(1\!+\!\mu_{k})|\mathbf{h}_{k}^{H}\mathbf{F}_{\mathrm{RF}}\mathbf{f}_{\mathrm{BB},k}|^{2}}{\sum_{j=1}^{K}\!|\mathbf{h}_{k}^{H}\mathbf{F}_{\mathrm{RF}}\mathbf{f}_{\mathrm{BB},j}|^{2}\!+\!\sigma_{k}^{2}},\!
\end{equation}
which is equivalent to the original objective function \eqref{eq:obj_a} when the auxiliary variables $\{\mu_{k}\}_{k=1}^{K}$ have the optimal value
\begin{equation}
\label{eq:mu}
\mu_{k}^{\star}=\frac{|\mathbf{h}_{k}^{H}\mathbf{F}_{\mathrm{RF}}\mathbf{f}_{\mathrm{BB},k}|^{2}}{\sum^{K}_{j=1,j\neq k}|\mathbf{h}_{k}^{H}\mathbf{F}_{\mathrm{RF}}\mathbf{f}_{\mathrm{BB},j}|^{2}+\sigma_{k}^{2}},~ \forall k.
\end{equation}
However, the summation of fractional terms in (\ref{eq:sum log}) still hinders a straightforward solution. Thus, we further apply the quadratic transform to convert it into
\begin{equation} \label{eq:quadratic transform}
2\sqrt{1+\mu_{k}}\Re\{\xi_{k}^{\ast}\mathbf{h}_{k}^{H}\mathbf{F}_{\mathrm{RF}}\mathbf{f}_{\mathrm{BB},k}\}-|\xi_{k}|^{2}C_{k}, ~\forall k,
\end{equation}
where $\{\xi_{k}\}_{k=1}^{K}$ are auxiliary variables. The expression (\ref{eq:quadratic transform}) is equivalent to the last fractional term in \eqref{eq:sum log} when the auxiliary variables $\{\xi_{k}\}_{k=1}^{K}$ have the optimal value
\begin{equation}
\label{eq:t}
\xi_{k}^{\star}=\frac{\sqrt{1+\mu_{k}}\mathbf{h}_{k}^{H}\mathbf{F}_{\mathrm{RF}}\mathbf{f}_{\mathrm{BB},k}}{C_{k}}, ~\forall k,
\end{equation}
where $C_{k}\triangleq \sum_{j=1}^{K}|\mathbf{h}_{k}^{H}\mathbf{F}_{\mathrm{RF}}\mathbf{f}_{\mathrm{BB},j}|^{2}+\sigma_{k}^{2}, \forall k$ is utilized to simplify the expression.

Based on the above derivation, the objective function can be reformulated as
\begin{equation} \label{eq: reformulated objective}
\begin{split}
&{\sum_{k=1}^{K}}\log_{2}(1+\mu_{k})-\mu_{k}\\
&\qquad+(2\sqrt{1+\mu_{k}}\Re\{\xi_{k}^{\ast}\mathbf{h}_{k}^{H}\!\mathbf{F}_{\mathrm{RF}}\mathbf{f}_{\mathrm{BB},k}\}-|\xi_{k}|^{2}C_{k}).
\end{split}
\end{equation}
In order to facilitate the subsequent transformation, (\ref{eq: reformulated objective}) can be rewritten as the following concise form
\begin{equation} \label{eq:concise form}
\sum_{k=1}^{K}(\log_{2}(1+\mu_{k})-\mu_{k}-|\xi_{k}|^{2}\sigma_{k}^{2})+\delta,
\end{equation}
where we define
\begin{equation} \label{eq:delta}
\begin{split}
\delta\triangleq & \sum_{k=1}^{K}(2\sqrt{1+\mu_{k}}\Re\{\xi_{k}^{\ast}\mathbf{h}_{k}^{H}\mathbf{F}_{\mathrm{RF}}\mathbf{f}_{\mathrm{BB},k}\}\\
& \hspace{1 cm}-\mathbf{f}_{\mathrm{BB},k}^{H}\mathbf{F}_{\mathrm{RF}}^{H}\mathbf{D}\mathbf{F}_{\mathrm{RF}}\mathbf{f}_{\mathrm{BB},k}),
\end{split}
\end{equation}
with $\mathbf{D}\triangleq \sum_{j=1}^{K}|\xi_{j}|^{2}\mathbf{h}_{j}\mathbf{h}_{j}^{H}$. After obtaining the auxiliary variables $\{\mu_{k}, \xi_{k}\}_{k=1}^{K}$ with the optimal value, maximizing (\ref{eq:concise form}) is equivalent to maximizing $\delta$.
Therefore, we can recast the optimization problem as
\begin{subequations}
\label{eq:transformed objective}
\begin{align}
    \max_{\mathbf{F}_{\mathrm{RF}}, \mathbf{F}_{\mathrm{BB}}}&\ \ \delta\\
    \label{eq:transformed objective b}
     \mathrm{s.t.}\ \ &\mathbf{F}_{\mathrm{RF}}(n,l)\in \{\mathcal{F},0\}, ~\forall n,l,\\
    \label{eq:transformed objective c}
    &\|\mathbf{F}_{\mathrm{RF}}(n,:)\|_{0}\in\{1,0\}, ~\forall n,\\
    \label{eq:transformed objective d}
    &\|\mathbf{F}_{\mathrm{RF}}\mathbf{F}_{\mathrm{BB}}\|_{F}^{2}= P_{\mathrm{t}}.
\end{align}
\end{subequations}
In the next two subsections, we endeavor to iteratively design the conditionally optimal analog beamformer $\mathbf{F}_{\mathrm{RF}}$ and digital beamformer $\mathbf{F}_{\mathrm{BB}}$ to address the problem \eqref{eq:transformed objective}.

\subsection{Dynamic Analog Beamformer Design}
The analog beamformer contains the information of phase-shift with constant modulus constraint \eqref{eq:transformed objective b} and subarray structure with $l_{0}$-norm constraint \eqref{eq:transformed objective c}, which cause obstructions for analog beamforming design. To tackle them, we take a divide-and-conquer approach. Specifically, the analog beamformer matrix $\mathbf{F}_{\mathrm{RF}}$ can be decomposed into two parts
\begin{equation}
    \label{eq:F=AS}
    \mathbf{F}_{\mathrm{RF}}=\mathbf{A}\mathbf{S},
\end{equation}
where $\mathbf{A}=\mathrm{diag}\{\pmb{\phi}\}\in \mathbb{C}^{N_{\mathrm{t}}\times N_{\mathrm{t}}}$ is a diagonal matrix including the information of phase-shift, where $\pmb{\phi}\triangleq[\phi_{1}, \phi_2, \ldots, \phi_{N_{\mathrm{t}}}]^{T}$ with the constant modulus constraints of $|\phi_{n}|={1}/{\sqrt{N_{\mathrm{t}}}},~\forall n$, and $\mathbf{S}\in \{0,1\}^{N_{\mathrm{t}}\times N_{\mathrm{RF}}}$ is a binary matrix representing the adjustable selection between array elements and the RF chains. After such decomposition, $\delta$ can be rewritten as
\begin{equation} \label{eq:delta 2}
\begin{split}
\delta  = & \sum_{k=1}^{K}\big(2\sqrt{1+\mu_{k}}\Re\{ \xi_{k}^{\ast}\mathbf{h}_{k}^{H}\mathbf{A}\mathbf{S}\mathbf{f}_{\mathrm{BB},k}\}  \\
& \hspace{1 cm}-\mathbf{f}_{\mathrm{BB},k}^{H}\mathbf{S}^{H}\mathbf{A}^{H} \mathbf{D}\mathbf{A}\mathbf{S}\mathbf{f}_{\mathrm{BB},k}\big).
\end{split}
\end{equation}

To sum up, when the auxiliary variables $\{\mu_{k}, \xi_{k}\}_{k=1}^{K}$ and digital beamformer matrix $\mathbf{F}_{\mathrm{BB}}$ are all fixed, the optimization problem \eqref{eq:transformed objective} for designing the analog beamformer $\mathbf{F}_{\mathrm{RF}}$ can be equivalently formulated as
\begin{subequations} \label{eq:A and S}
\begin{align}
    \max_{\mathbf{A}, \mathbf{S}}\ \ &\delta\\
    \mathrm{s.t.}\ \ &\mathbf{A}=\mathrm{diag}\{\pmb{\phi}\}, \\
    &|\phi_{n}|={1}/{\sqrt{N_{\mathrm{t}}}}, ~\forall n,\\
    &\mathbf{S}(n,l)\in\{1,0\}, ~\forall n,l,\\
    &\|\mathbf{S}(n,:)\|_{0}\in\{1,0\}, ~\forall n,
\end{align}
\end{subequations}
where the power constraint \eqref{eq:transformed objective d} is disregarded due to the fact that the power can be adjusted by the digital beamformer. In the following, we will employ the binary integer programming approach to calculate the switch matrix $\mathbf{S}$, and utilize the manifold algorithm to obtain the phase-shift matrix $\mathbf{A}$.

\subsubsection{Switch Matrix $\mathbf{S}$}
For (\ref{eq:A and S}), with the fixed phase-shift matrix $\mathbf{A}$, it can be simplified to a typical 0/1 integer programming problem as:
\begin{subequations}
\label{eq:S}
\begin{align}
    \max_{\mathbf{S}}\ \ &\delta\\
    \mathrm{s.t.}\ \ &\mathbf{S}(n,l)\in\{1,0\}, ~\forall n,l,\\
    &\|\mathbf{S}(n,:)\|_{0}\in\{1,0\}, ~\forall n,
\end{align}
\end{subequations}
where the optimal $\mathbf{S}^{\star}$ can be found by some off-the-shelf solvers (e.g., Mosek optimization tools).

\subsubsection{Phase-shift Matrix $\mathbf{A}$}
After obtaining the switch matrix $\mathbf{S}$, the optimization problem of phase-shift matrix $\mathbf{A}$ can be expressed as
\begin{subequations} \label{eq: optimizing A}
\begin{align}
    \max_{\mathbf{A}}\ \ &\delta\\
    \mathrm{s.t.}\ \ &\mathbf{A}=\mathrm{diag}\{\pmb{\phi}\}, \\
    \label{eq: optimizing A c}
    &|\phi_{n}|=1/\sqrt{N_\text{t}}, ~\forall n,
\end{align}
\end{subequations}
where \eqref{eq: optimizing A c} is a non-convex constraint that causes significant difficulties in solving the problem. Considering the performance and effectiveness of various existing algorithms, we employ the manifold-based algorithm in this paper. First, by utilizing the equation $\mathbf{A}\mathbf{S}\mathbf{f}_{\mathrm{BB},k}=\mathrm{diag}\{\mathbf{S}\mathbf{f}_{\mathrm{BB},k}\}\pmb{\phi}$, $\delta$ in (\ref{eq:delta 2}) can be recast as
\begin{equation} \label{eq:delta 3}
\begin{split}
\delta= & \sum_{k=1}^{K} \big(2\sqrt{1+\mu_{k}}\Re\{\xi_{k}^{\ast}\mathbf{h}_{k}^{H}\mathrm{diag}\{\mathbf{S}\mathbf{f}_{\mathrm{BB},k}\}\pmb{\phi}\}\\
& \hspace{1 cm}-\pmb{\phi}^{H}\mathrm{diag}\{\mathbf{f}_{\mathrm{BB},k}^{H}\mathbf{S}^{H}\}\mathbf{D}\mathrm{diag}\{\mathbf{S}\mathbf{f}_{\mathrm{BB},k}\}\pmb{\phi} \big).
\end{split}
\end{equation}
After that, problem (\ref{eq: optimizing A}) can be re-arranged as
\begin{subequations}
\label{eq: optimizing A_2}
\begin{align}
    \label{eq: optimizing A_2 a}
    \min_{\pmb{\phi}}\ \ &\pmb{\phi}^{H}\mathbf{Q}\pmb{\phi}-2\Re\{\pmb{\phi}^{H}\mathbf{q}\},\\
    \label{eq: optimizing A_2 b}
    \mathrm{s.t.}\ \ &|\phi_{n}|=1/\sqrt{N_\text{t}}, ~\forall n,
\end{align}
\end{subequations}
where for brevity we define
\begin{subequations}\begin{align}
\mathbf{Q}&\triangleq \sum_{k=1}^{K}\mathrm{diag}\{\mathbf{f}_{\mathrm{BB},k}^{H}\mathbf{S}^{H}\}\mathbf{D}\mathrm{diag}\{\mathbf{S}\mathbf{f}_{\mathrm{BB},k}\},\\
\mathbf{q}&\triangleq \sum_{k=1}^{K}\sqrt{1+\mu_{k}}\xi_{k}\mathrm{diag}\{\mathbf{f}_{\mathrm{BB},k}^{H}\mathbf{S}^{H}\}\mathbf{h}_{k}.
\end{align}\end{subequations}
Then, we observe that the objective function \eqref{eq: optimizing A_2 a} is smooth with easy-to-obtain derivatives. Additionally, the unit-modulus constraint \eqref{eq: optimizing A_2 b} forms a complex circle Riemannian manifold, which allows problem     \eqref{eq: optimizing A_2} to be solved by the typical Riemannian conjugate gradient (RCG) algorithm. After deriving the Riemannian gradient from the corresponding Euclidean gradient, it can be iteratively solved on the Riemannian space by utilizing the idea of the conjugate gradient algorithm. Readers can refer to \cite{R. Liu 2021} for more details. After having the phase-shift matrix $\mathbf{A}$ and the switch matrix $\mathbf{S}$, the analog beamforming matrix $\mathbf{F}_{\mathrm{RF}}$ can be constructed by \eqref{eq:F=AS}.

\vspace{-0.2cm}
\subsection{Digital Beamformer Design}
With given auxiliary variables $\{\mu_{k}, \xi_{k}\}_{k=1}^{K}$, and the analog beamforming matrix $\mathbf{F}_{\mathrm{RF}}$, the digital beamforming matrix $\mathbf{F}_{\mathrm{BB}}$ is updated by solving the following optimization problem
\vspace{-0.4 cm}
\begin{subequations}\label{eq:digital beamformer}
\begin{align}
\max_{\mathbf{F}_{\mathrm{BB}}}&\ \ \delta\\
\label{eq:digital beamformer b}
\mathrm{s.t.}
&\ \ \|\mathbf{F}_{\mathrm{RF}}\mathbf{F}_{\mathrm{BB}}\|_{F}^{2}= P_{\mathrm{t}}.
\end{align}
\end{subequations}
If the transmit power constraint \eqref{eq:digital beamformer b} is temporarily disregarded, each column of $\mathbf{F}_{\mathrm{BB}}$ can be optimized separately by letting the partial derivative w.r.t. $\mathbf{f}_{\mathrm{BB},k}$ of the objective in \eqref{eq:delta} to zero, which yields
\begin{equation}
\label{eq:digital beamformer result}
\mathbf{\overline{f}}_{\mathrm{BB},k} = (\mathbf{F}_{\mathrm{RF}}^{H}\mathbf{D}\mathbf{F}_{\mathrm{RF}})^{-1}\sqrt{(1+\mu_{k})}\xi_{k}\mathbf{F}_{\mathrm{RF}}^{H}\mathbf{h}_{k},~ \forall k.
\end{equation}
After obtaining the semi-finished digital beamformer $\mathbf{\overline{F}}_{\mathrm{BB}}=[\mathbf{\overline{f}}_{\mathrm{BB},1}, \mathbf{\overline{f}}_{\mathrm{BB},2}, \ldots, \mathbf{\overline{f}}_{\mathrm{BB},K}]$, we need to scale it to make it satisfy the total power constraint, which can be calculated as
\begin{equation}
    \label{eq:digital beamformer power}
    \mathbf{F}_{\mathrm{BB}}^{\star}=\frac{\sqrt{P_{\mathrm{t}}}\ \mathbf{\overline{F}}_{\mathrm{BB}}}{\|\mathbf{F}_{\mathrm{RF}}\mathbf{\overline{F}}_{\mathrm{BB}}\|_{F}}.
\end{equation}

\begin{algorithm}[!t]
\caption{The proposed real-time dynamic hybrid beamforming design.}
\label{alg:1}
    \begin{algorithmic}[1]
    \begin{small}
    \REQUIRE $\{\mathbf{h}_{k}\}_{k=1}^{K}$, $\{\sigma_k^2\}_{k=1}^{K}$, and $P_\text{t}$.
    \ENSURE $\mathbf{F}_{\mathrm{RF}}^{\star}$, $\mathbf{F}_{\mathrm{BB}}^{\star}$.
            \STATE {Initialize $\mathbf{F}_{\mathrm{RF}}$ and $\mathbf{F}_{\mathrm{BB}}$.}
            \WHILE {no convergence of $\mathbf{F}_{\mathrm{RF}}$ and $\mathbf{F}_{\mathrm{BB}}$}
                   \STATE {Update $\{\mu_{k}^{\star}\}_{k=1}^{K}$ by \eqref{eq:mu}.}
                   \STATE {Update $\{\xi_{k}^{\star}\}_{k=1}^{K}$ by \eqref{eq:t}.}
                   \STATE {Update the switch matrix $\mathbf{S}^{\star}$ by solving  \eqref{eq:S}.}
                   \STATE {Update the phase-shift matrix $\mathbf{A}^{\star}$ by solving \eqref{eq: optimizing A_2}.}
                   \STATE {Update the analog beamforming matrix $\mathbf{F}_{\mathrm{RF}}^{\star}$ by \eqref{eq:F=AS}.}
                   \STATE {Update the digital beamforming matrix $\mathbf{F}_{\mathrm{BB}}^{\star}$ by \eqref{eq:digital beamformer result} and \eqref{eq:digital beamformer power}.}
            \ENDWHILE
            \STATE {Return $\mathbf{F}_{\mathrm{RF}}^{\star}$ and $\mathbf{F}_{\mathrm{BB}}^{\star}$.}
    \end{small}
    \end{algorithmic}
\end{algorithm}

\begin{figure*}[!t]
  \centering
  \includegraphics[width= 0.92\linewidth]{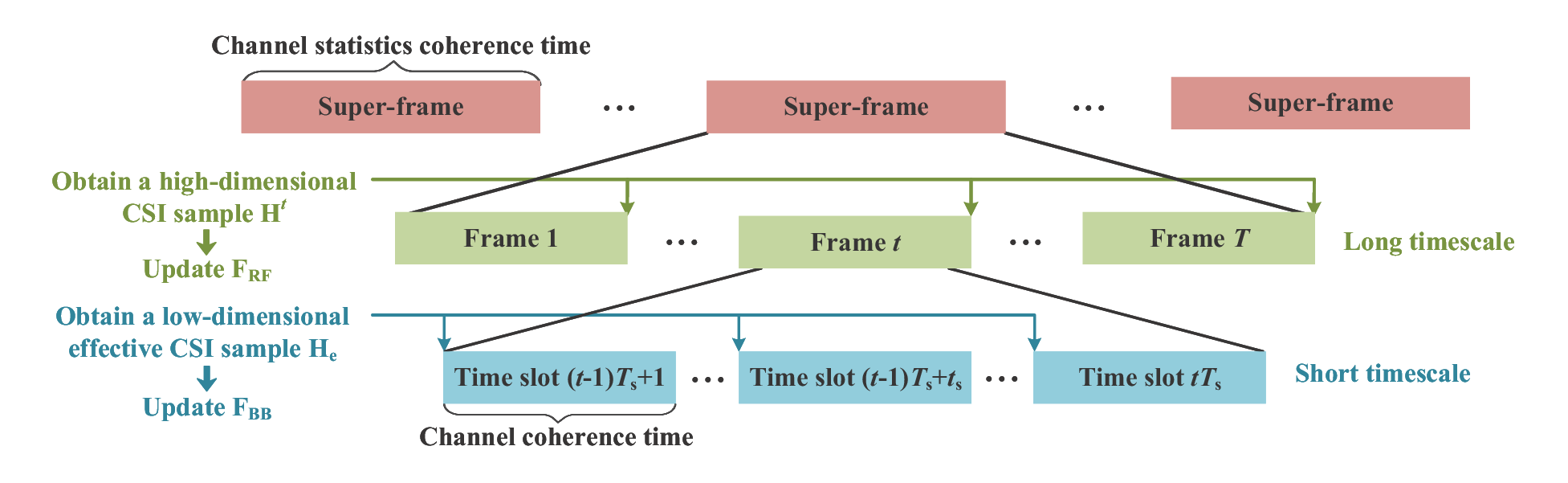}
  \caption{ The frame structure of two-timescale dynamic hybrid beamforming.}
  \label{fig:S-CSI}
\end{figure*}

\subsection{Summary, Complexity and Convergence Analysis}
Based on the above derivations, the proposed FP-manifold-based real-time dynamic hybrid beamforming design is straightforward and summarized in Algorithm 1. After initializing $\mathbf{F}_{\mathrm{RF}}$ and $\mathbf{F}_{\mathrm{BB}}$ appropriately, the auxiliary variables $\{\mu_{k}, \xi_{k}\}_{k=1}^{K}$, the switch matrix $\mathbf{S}$, the phase-shift matrix $\mathbf{A}$, and the digital beamformer $\mathbf{F}_{\mathrm{BB}}$ are iteratively updated until a convergence solution is obtained.

Then we briefly analyze the algorithmic complexity. Specifically, computing the switch matrix $\mathbf{S}$ in \eqref{eq:S} needs to solve an $N_{\mathrm{RF}}$-dimensional integer programming optimization problem with $N_{\mathrm{t}}$ variables by CVX, which has $\mathcal{O}\{\sqrt{N_{\mathrm{t}}N_{\mathrm{RF}}}[N_{\mathrm{RF}}^{3}N_{\mathrm{t}}^{2}+N_{\mathrm{RF}}^{2}N_{\mathrm{t}}^{3}]\}$ complexity \cite{Lectures 2001}. Besides, the complexity to obtain $\mathbf{A}$ is at most $\mathcal{O}\{N_{\mathrm{t}}^{1.5}\}$ using the RCG algorithm in \eqref{eq: optimizing A_2}. The complexity of calculating the digital beamformer in  \eqref{eq:digital beamformer result} is $\mathcal{O}\{KN_{\mathrm{RF}}N_{\mathrm{t}}^{2}\}$. As a result, the overall complexity of the proposed algorithm is $\mathcal{O}\{N_{\mathrm{iter}}(KN_{\mathrm{RF}}N_{\mathrm{t}}^{2}+N_{\mathrm{t}}^{1.5}+N_{\mathrm{RF}}^{3.5}N_{\mathrm{t}}^{2.5}+N_{\mathrm{RF}}^{2.5}N_{\mathrm{t}}^{3.5})\}$, where $N_{\mathrm{iter}}$ is the number of iterations.

Regarding the convergence, since framework of the real-time dynamic hybrid beamforming design is similar to
\cite{H. Li 2020 Hybrid}, the convergence of the proposed algorithm can be proved by the same procedure offered in \cite{H. Li 2020 Hybrid}. Due to space limitations, the detailed derivation is omitted here. Please refer to \cite{H. Li 2020 Hybrid} for a comprehensive explanation.

\vspace{0.5cm}
\section{Two-timescale Dynamic Hybrid Beamforming Design}
When the instantaneous high-dimensional CSI cannot be easily acquired due to significant channel estimation overhead, the proposed FP-manifold-based real-time dynamic hybrid beamforming design cannot be directly applied. To realize an efficient hybrid beamforming design in such practical scenarios, we utilize a two-timescale hybrid beamforming framework, which adapts the analog beamformer in the long-timescale based on the statistical high-dimensional channel samples and designs the digital beamformer in the short-timescale with instantaneous low-dimensional effective CSI. In the following, we will first provide a brief overview of the two-timescale framework and then elaborate on its detailed algorithmic implementation for tackling the dynamic hybrid beamforming design challenge.

\subsection{Two-timescale Hybrid Beamforming Framework}
The frame structure of the two-timescale algorithm is illustrated in Fig. \ref{fig:S-CSI}.
In a detailed explanation, the time domain is partitioned into several super-frames referred to as channel statistics coherence time, during which the channel statistical distribution remains consistent. Each super-frame is then further divided into $T$ frames, each consisting of $T_{\mathrm{s}}$ time-slots termed channel coherence time, during which the channel state is assumed constant. In accordance with this structure, we can define the following two timescales.
\begin{itemize}
  \item Long-timescale: The BS acquires one statistical high-dimensional channel sample at the end of each frame for the long-timescale analog beamformer design to achieve the array gain.
  \item Short-timescale: The BS obtains one instantaneous low-dimensional effective CSI at the start of each time-slot for short-timescale digital beamformer design to achieve the spatial multiplexing gain.
\end{itemize}
It can be observed that only a few high-dimensional channel samples are required in each super-frame and the low-dimensional effective CSI can be efficiently obtained, significantly reducing channel estimation overhead.
Besides, in contrast to conventional statistical-CSI-based beamforming designs which rely on the explicit knowledge of the channel statistics, this online approach learns and updates the channel statistical information implicitly by observing one channel sample at each frame. Thus, it requires fewer channel measurement samples and less memory, leading to reduced latency.
Moreover, as the method for acquiring statistical channel information remains the same for various types of channels, the two-timescale algorithm is adaptable to both far-field and near-field cases,  highlighting its universality and broad applicability.

Overall, this two-timescale algorithm can reduce the training overhead and computational complexity associated with channel estimation, simplify the process of dynamically adjusting the switch network by designing the analog beamformer in the long-timescale, and ensure satisfactory performance by designing the digital beamformer in the short-timescale. In the following subsections, we will describe the dynamic hybrid beamforming design procedure based on this framework.

\subsection{Long-timescale Analog Beamformer Design}
In this subsection, we aim to solve the long-timescale analog beamforming design problem with fixed digital beamformer $\mathbf{F}_{\mathrm{BB}}$, which can be written as
\begin{subequations}
\label{eq:obj_analog}
\begin{align}
    \label{eq:obj_analog_a}
    \max_{\mathbf{F}_{\mathrm{RF}}}&\ \ \sum^{K}_{k=1}\mathbb{E}_{\mathbf{H}}\left\{\log_{2}(1+\mathrm{SINR}_{k})\right\}\\
    \label{eq:obj_analog_b}
    \mathrm{s.t.}&\ \ \mathbf{F}_{\mathrm{RF}}(n,l)\in \{\mathcal{F},0\},~ \forall n,l,\\
    \label{eq:obj_analog_c}
    &\ \ \|\mathbf{F}_{\mathrm{RF}}(n,:)\|_{0}\in\{1,0\},~\forall n,
\end{align}
\end{subequations}
where the ergodic achievable sum-rate of near-field communication over channel $\mathbf{H}\triangleq[\mathbf{h}_{1}, \mathbf{h}_{2}, \ldots, \mathbf{h}_{K}]\in \mathbb{C}^{N_{\mathrm{t}}\times K}$ should be taken as the objective function with the same constraints as the real-time circumstance. The difficulties of resolving this problem lie in the stochastic objective \eqref{eq:obj_analog_a} and the non-smooth and non-convex Boolean constraints \eqref{eq:obj_analog_b}, \eqref{eq:obj_analog_c}. In the following, we first substitute the expectation in the objective function with a quadratic surrogate function according to the SSCA algorithm \cite{A. Liu two-stage}. Then, we convert the Boolean constraints into quadratic penalty terms with a box constraint and employ the MM method and smooth logarithmic approximation technique to make them more tractable. Thus, the analog beamformer can be obtained at the end of each frame by iteratively solving a convex problem.

\subsubsection{Objective Function Substitution}
In accordance with the SSCA algorithm described in \cite{A. Liu two-stage}, the premise for utilizing the surrogate function to substitute the objective function \eqref{eq:obj_analog_a} is that the variable should be real-valued and continuous. To satisfy this prerequisite and facilitate the subsequent algorithm development, the analog beamformer matrix should also be divided into two halves, i.e.,   $\mathbf{F}_{\mathrm{RF}}=\mathbf{A}\mathbf{S}$ where $\mathbf{A}=\mathrm{diag}\{\pmb{\phi}\}\in \mathbb{C}^{N_{\mathrm{t}}\times N_{\mathrm{t}}}$ contains the phase-shift information and $\mathbf{S}\in \{0,1\}^{N_{\mathrm{t}}\times N_{\mathrm{RF}}}$ represents the switch matrix, corresponding to \eqref{eq:F=AS}. The optimization problem can be recast as
\begin{subequations}
\label{eq:obj_analog_AS}
\begin{align}
    \label{eq:obj_analog_AS_a}
    \max_{\mathbf{A},\mathbf{S}}&\ \ \sum^{K}_{k=1}\mathbb{E}_{\mathbf{H}}\left\{\log_{2}(1+\mathrm{SINR}_{k})\right\}\\
    \label{eq:obj_analog_AS_b}
    \mathrm{s.t.}&\ \ \mathbf{S}(n,l)\in \{1,0\}, ~\forall n,l,\\
    \label{eq:obj_analog_AS_c}
    &\ \ \|\mathbf{S}(n,:)\|_{0}\in\{1,0\}, ~\forall n.
\end{align}
\end{subequations}
Then, the real-valued angle vector $\bm{\theta}\in \mathbb{R}^{N_{\mathrm{t}}\times 1}$ can be extracted from the complex valued phase-shift vector $\pmb{\phi}$, i.e., $\bm{\theta}=\angle\pmb{\phi}$. Moreover, the switch matrix $\mathbf{S}$ can be converted into a vector as $\mathbf{s} \triangleq \mathrm{vec}(\mathbf{S})\in \{0,1\}^{N_{\mathrm{t}}N_{\mathrm{RF}}\times 1}$. We denote $s_{n,l}$ as the $(n+(l-1)N_{\mathrm{t}})$-th element of switch vector $\mathbf{s}$, which is also equal to $\mathbf{S}(n,l)$.
Thus, the $n$-th row of $\mathbf{S}$ can be equivalently represented as
\begin{equation}\label{S convert to s}
\begin{split}
\mathbf{S}(n,:)\triangleq[s_{n,1}, s_{n,2}, \ldots, s_{n,N_{\mathrm{RF}}}]\in \{0,1\}^{1\times N_{\mathrm{RF}}}, ~\forall n.
\end{split}
\end{equation}
With newly introduced variable $\bm{\theta}$ and $\mathbf{s}$, the problem \eqref{eq:obj_analog_AS} can be reformulated as
\begin{subequations}
\label{eq:objTOSCA}
\begin{align}
    \label{eq:objTOSCA_a}
    \min_{\bm{\theta},\mathbf{s}}&\ \ \mathbb{E}_{\mathbf{H}}\{g_{0}(\bm{\theta}, \mathbf{s}; \mathbf{H})\}\\
    \mathrm{s.t. }\label{eq:objTOSCA_b} &\ \ s_{n,l}\in\{1,0\}, ~\forall n,l,\\
                  \label{eq:objTOSCA_c} &\ \ \left\|[s_{n,1}, s_{n,2}, \ldots, s_{n,N_{\mathrm{RF}}}]\right\|_{0}\in\{1,0\},~\forall n,
\end{align}
\end{subequations}
where the objective function is defined as $\mathbb{E}_{\mathbf{H}}\{g_{0}(\bm{\theta}, \mathbf{s}; \mathbf{H})\}\triangleq -\mathbb{E}_{\mathbf{H}}\{\sum_{k=1}^{K}\log_{2}(1+\mathrm{SINR}_{k})\}$ for brevity.

In order to tackle the difficulty that no closed-form expression of $\mathbb{E}_{\mathbf{H}}\{g_{0}(\bm{\theta}, \mathbf{s}; \mathbf{H})\}$ is available, we construct a convex surrogate quadratic function to substitute it based on the channel samples obtained at the current and previous frames. Then, by continuous sampling, the BS can recursively learn the statistical properties of the channel. This accumulation of channel samples gradually improves the beamforming performance and eventually achieves convergence.

Inspired by the well-known SSCA algorithm, we can use statistical samples to recursively minimize a sequence of surrogate functions that must be convex but not necessarily an upper bound of the objective function (\ref{eq:objTOSCA_a}). Specifically, in the $t$-th recursion, based on the channel sample $\mathbf{H}^{t}$ obtained at the end of $t$-th frame and the current analog beamformer information (i.e., $\bm{\theta}^{t}$ and $\mathbf{s}^{t}$) used in this frame, we can replace $\mathbb{E}_{\mathbf{H}}\{g_{0}( \bm{\theta},\mathbf{s}; \mathbf{H})\}$ with a quadratic surrogate function $f^{t}(\bm{\theta}, \mathbf{s})$ as
\begin{equation}\label{surrogate}
\begin{split}
   f^{t}(\bm{\theta}, \mathbf{s})= v^{t} &+ (\mathbf{v}_{\theta}^{t})^{T}(\bm{\theta}-\bm{\theta}^{t}) + (\mathbf{v}_{s}^{t})^{T}(\mathbf{s}-\mathbf{s}^{t})\\
   &+ \tau\|\bm{\theta}-\bm{\theta}^{t}\|^{2}+ \tau\|\mathbf{s}-\mathbf{s}^{t}\|^{2},
\end{split}
\end{equation}
where $\tau>0$ can be any constant, the terms $\tau\|\bm{\theta}-\bm{\theta}^{t}\|^{2}$ and $\tau\|\mathbf{s}-\mathbf{s}^{t}\|^{2}$ are used to ensure strong convexity, $\rho^{t}\in(0,1]$ is a properly chosen sequence to satisfy the convergence requirement,
and $v^{t}$ is the approximation of the objective function value $\mathbb{E}_{\mathbf{H}}\{g_{0}(\bm{\theta}^{t}, \mathbf{s}^{t}; \mathbf{H}^{t})\}$, which can be calculated recursively as
\begin{equation}\label{constant}
   v^{t} = (1-\rho^{t})v^{t-1} + \rho^{t}g_{0}(\bm{\theta}^{t}, \mathbf{s}^{t}; \mathbf{H}^{t}).
\end{equation}
Vectors $\mathbf{v}_{\theta}^{t}$ and $\mathbf{v}_{s}^{t}$ are the approximations of the partial derivative of $\mathbb{E}_{\mathbf{H}}\{g_{0}(\bm{\theta}, \mathbf{s}; \mathbf{H})\}$ w.r.t. $\bm{\theta}$ at $\bm{\theta}=\bm{\theta}^{t}$ and w.r.t. $\mathbf{s}$ at $\mathbf{s}=\mathbf{s}^{t}$, respectively, which can be calculated recursively as
\begin{subequations}\begin{align}\label{surrogate_gradient_theta}
   \mathbf{v}_{\theta}^{t} &= (1-\rho^{t})\mathbf{v}_{\theta}^{t-1} + \rho^{t}\nabla_{\bm{\theta}}g_{0}(\bm{\theta}^{t}, \mathbf{s}^{t}; \mathbf{H}^{t}),\\
\label{surrogate_gradient_s}
   \mathbf{v}_{s}^{t} &= (1-\rho^{t})\mathbf{v}_{s}^{t-1} + \rho^{t}\nabla_{\mathbf{s}}g_{0}(\bm{\theta}^{t}, \mathbf{s}^{t}; \mathbf{H}^{t}).
\end{align}\end{subequations}
Then, the gradients of $g_{0}(\bm{\theta},\mathbf{s}; \mathbf{H})$ w.r.t. $\bm{\theta}$ and w.r.t. $\mathbf{s}$ in \eqref{surrogate_gradient_theta} and \eqref{surrogate_gradient_s} are given by
\begin{subequations}\begin{align}\label{gradient_theta}
   \nabla_{\bm{\theta}}g_{0}(\bm{\theta},\mathbf{s}; \mathbf{H}) &= -\sum_{k=1}^{K}\left(\frac{\sum_{i}^{K}\mathbf{a}^{\theta}_{k,i}}{\mit{\Gamma_{k}}}-\frac{\sum_{i\neq k}^{K}\mathbf{a}^{\theta}_{k,i}}{\mit{\Gamma_{-k}}}\right),\\
\label{gradient_s}
   \nabla_{\mathbf{s}}g_{0}(\bm{\theta},\mathbf{s}; \mathbf{H}) &= -\sum_{k=1}^{K}\left(\frac{\sum_{i}^{K}\mathbf{a}^{s}_{k,i}}{\mit{\Gamma_{k}}}-\frac{\sum_{i\neq k}^{K}\mathbf{a}^{s}_{k,i}}{\mit{\Gamma_{-k}}}\right),
\end{align}\end{subequations}
where we define
\begin{subequations}\begin{align}
\mit{\Gamma_{k}} \!&=  \! \sum_{i}^{K}\left|\mathbf{h}_{k}^{H}\mathbf{F}_{\mathrm{RF}}\mathbf{f}_{\mathrm{BB},i}\right|^{2}+\sigma_{k}^{2}, ~\forall k, \\
\mit{\Gamma_{-k}}\! &=  \!\sum_{i\neq k}^{K}\left|\mathbf{h}_{k}^{H}\mathbf{F}_{\mathrm{RF}}\mathbf{f}_{\mathrm{BB},i}\right|^{2}+\sigma_{k}^{2}, ~\forall k,\\
\mathbf{a}^{\theta}_{k,i}\!  &= \!-2\Re\{\jmath\pmb{\phi}^{\ast}\odot(\mathbf{Q}_{k,i}^{H}\pmb{\phi})\}, ~\forall k, i,\\
\mathbf{a}^{s}_{k,i} \!&=  \! 2\Re\{(\mathbf{I}_{K}\otimes\mathbf{A}^{H})\mathrm{vec}(\mathbf{h}_{k}\mathbf{h}_{k}^{H}\!\!\mathbf{A}\mathbf{S}\mathbf{f}_{\mathrm{BB},i}\mathbf{f}_{\mathrm{BB},i}^{H})\}, ~\forall k,i,\!\!
\end{align}\end{subequations}
with $\mathbf{Q}_{k,i}=\mathrm{diag}^{H}\{\mathbf{S}\mathbf{f}_{\mathrm{BB},i}\}\mathbf{h}_{k}\mathbf{h}_{k}^{H}\mathrm{diag}\{\mathbf{S}\mathbf{f}_{\mathrm{BB},i}\}$. The detailed derivations are omitted due to space limitations.

Afterward, $f^{t}(\bm{\theta},\mathbf{s})$ can be viewed as a convex approximation of the original stochastic non-convex objective function $\mathbb{E}_{\mathbf{H}}\{g_{0}(\bm{\theta},\mathbf{s}; \mathbf{H})\}$, which allows us to handle it without calculating the expectation explicitly. Therefore, at the end of the $t$-th frame, with the channel sample $\mathbf{H}^{t}$ and the current analog beamformer information (i.e., $\bm{\theta}^{t}$ and $\mathbf{s}^{t}$), we need to solve the following optimization problem
\begin{subequations}
\label{eq:objTOSCAsurrogate_a}
\begin{align}
   \!\!\!(\bm{\bar{\theta}}^{t}, \mathbf{\bar{s}}^{t})\!=& \arg\min_{\bm{\theta},\mathbf{s}} f^{t}(\bm{\theta},\mathbf{s})\\
   &\ \mathrm{s.t. }\ \label{eq:objTOSCAsurrogate_b} \ s_{n,l}\in\{1,0\},~ \forall n, l,\\
                  &\ \ \ \ \ \ \label{eq:objTOSCAsurrogate_c} \left\|[s_{n,1}, s_{n,2}, \ldots, s_{n,N_{\mathrm{RF}}}]\right\|_{0}\!\in\!\{1,0\},~\forall n,
\end{align}
\end{subequations}
 to obtain $\bm{\bar{\theta}}^{t}$ and $\mathbf{\bar{s}}^{t}$, and then utilize them to update the analog beamformer for the next frame.
\subsubsection{Tackle Boolean Constraints}
After dealing with the objective function, we turn to tackle the non-convex Boolean constraints in \eqref{eq:objTOSCAsurrogate_a}. Notice that there are two Boolean constraints, where \eqref{eq:objTOSCAsurrogate_b} indicates that the variable $\mathbf{s}$ is binary, and \eqref{eq:objTOSCAsurrogate_c} restricts the $l_{0}$-norm of the vector $[s_{n,1}, s_{n,2}, \ldots, s_{n,N_{\mathrm{RF}}}]\in\{0,1\}^{1\times N_{\mathrm{RF}}}$ to be binary. In addition, due to the property of $l_{0}$-norm, \eqref{eq:objTOSCAsurrogate_c} is neither smooth nor convex. Next, we will apply some transformations and approximations to handle these constraints.

\newcounter{TempEqCnt}
\setcounter{TempEqCnt}{\value{equation}}
\setcounter{equation}{47}
\begin{figure*}[b]
\hrulefill
\begin{equation}\label{eq:log_s_linear}
\begin{split}
   \tilde{s}_{n}\approx&\sum_{l=1}^{N_{\mathrm{RF}}}\frac{1}{\log(1+{1}/{\epsilon})}\Big( \log\big(1+{s_{n,l}^{(m)}}/{\epsilon} \big)+ \frac{1}{\epsilon+s_{n,l}^{(m)}}\big(s_{n,l} - s_{n,l}^{(m)}\big)\Big),~\forall n, s_{n,l}\in[0,1].
\end{split}
\end{equation}
\begin{equation}\label{eq:log_s_linear_matrix}
\mathbf{\tilde{s}}=\frac{1}{\log(1+{1}/{\epsilon})}\Big(\mathbf{T}\log\big(1+{\mathbf{s}^{(m)}}/{\epsilon}\big)+ \mathbf{T}\mathrm{diag}^{-1}\big\{\epsilon+\mathbf{s}^{(m)}\big\}\big(\mathbf{s}-\mathbf{s}^{(m)}\big)\Big)
=\mathbf{T}\big(\mathbf{u}^{(m)}+\mathbf{U}^{(m)}(\mathbf{s}-\mathbf{s}^{(m)})\big).
\end{equation}
\end{figure*}

\setcounter{equation}{40}
According to \cite{Boolean}, the Boolean constraint can be converted into solving the problem of minimizing a quadratic term with a box constraint, which is smooth and more tractable. To be specific, \eqref{eq:objTOSCAsurrogate_b} is equivalent to seeking the optimal solution for the following problem
\begin{subequations}
\label{eq:Boolean1_a}
\begin{align}
    \label{eq:Boolean1_b}
    \min_{\mathbf{s}}&\ \ \mathbf{s}^{T}(\mathbf{1}-\mathbf{s})\\
    \mathrm{s.t. }\label{eq:Boolean1_c} &\ \ s_{n,l}\in[0,1], ~\forall n,l,
\end{align}
\end{subequations}
where $s_{n,l}$, denoting the $(n+(l-1)N_{\mathrm{t}})$-th element of the switch vector $\mathbf{s}$, is relaxed to a continuous variable. Likewise, by introducing an auxiliary variable $\mathbf{\tilde{s}} \triangleq [\tilde{s}_{1}, \tilde{s}_{2}, \ldots, \tilde{s}_{N_{\mathrm{t}}}]^{T}\in\mathbb{R}^{N_{\mathrm{t}}\times 1}$ with $\tilde{s}_{n} \triangleq \left\|[s_{n,1}, s_{n,2}, \ldots, s_{n,N_{\mathrm{RF}}}]\right\|_{0}$, the constraint \eqref{eq:objTOSCAsurrogate_c} is equivalent to seeking the optimal variable $\mathbf{s}$ for the subsequent optimization problem
\begin{subequations}
\label{eq:Boolean2_a}
\begin{align}
    \label{eq:Boolean2_b}
    \min_{\mathbf{s}}&\ \ \mathbf{\tilde{s}}^{T}(\mathbf{1}-\mathbf{\tilde{s}})\\
    \mathrm{s.t. }\label{eq:Boolean2_c} &\ \ s_{n,l}\in[0,1], ~\forall n,l.
\end{align}
\end{subequations}
The above two problems \eqref{eq:Boolean1_a} and \eqref{eq:Boolean2_a} can be incorporated as penalty terms into the objective function with a box constraint, which results in a new optimization problem as
\begin{subequations}
\label{eq:obj_Boolean_a}
\begin{align}
   \label{eq:obj_Boolean_b}(\bm{\bar{\theta}}^{t}, \mathbf{\bar{s}}^{t})\!=& \arg\min_{\bm{\theta}, \mathbf{s}} f^{t}(\bm{\theta}, \mathbf{s}; \mathbf{H})+\varrho_{1}\mathbf{s}^{T}(\mathbf{1}-\mathbf{s}) + \varrho_{2}\mathbf{\tilde{s}}^{T}(\mathbf{\mathbf{1}-\tilde{s}})\\
   &\ \mathrm{s.t. }\label{eq:obj_Boolean_c}\ \ s_{n,l}\in[0,1], ~\forall n,l,
\end{align}
\end{subequations}
where $\varrho_{1}$ and $\varrho_{2}$ are the penalty parameters to control the trade-off between the value of the objective and the degree of satisfying the Boolean constraints.

\begin{algorithm}[!t]
\caption{The inner iteration of optimizing phase-shift vector and switch vector.}
\label{alg:2}
    \begin{algorithmic}[1]
    \begin{small}
    \REQUIRE $\mathbf{H}^{t}, \bm{\theta}^{t}, \mathbf{s}^{t}, \varrho_{1}, \varrho_{2}, N_{\mathrm{max}}$.
    \ENSURE $\bm{\bar{\theta}}^{t}$ and $\mathbf{\bar{s}}^{t}$ .
            \STATE {Set $m=1$, $\bm{\theta}^{(1)}=\bm{\theta}^{t}$, $\mathbf{s}^{(1)}=\mathbf{s}^{t}$.}
            \WHILE {$m<N_{\mathrm{max}}$ $\pmb{\&}$ no convergence of $\bm{\theta}^{(m)}$ and $\mathbf{s}^{(m)}$}
                   \STATE {Increase the index of iteration $m: =m+1$.}
                   \STATE {Update  $\bm{\theta}^{(m)}$ and  $\mathbf{s}^{(m)}$ by solving \eqref{eq:obj_final}.}
                   \STATE {Adjust the penalty parameters $\varrho_{1}:=c_{1}\varrho_{1}$, $ \varrho_{2}:=c_{2}\varrho_{2}$. }
            \ENDWHILE
            \STATE Let {$\bm{\bar{\theta}}^{t}=\bm{\theta}^{(m)}$ and $\mathbf{\bar{s}}^{t}=\mathbf{s}^{(m)}$.}
            \STATE {Return $\bm{\bar{\theta}}^{t}$ and $\mathbf{\bar{s}}^{t}$.}
    \end{small}
    \end{algorithmic}
\end{algorithm}

\begin{algorithm}[t]
\caption{The proposed two-timescale dynamic hybrid beamforming design.}
\label{alg:3}
    \begin{algorithmic}[0]
    \begin{small}
            \STATE {\!\!\!\!\!\!\textbf{Initialize} $\bm{\theta}^{1}, \mathbf{s}^{1}, \mathbf{v}_{\theta}^{0}=\mathbf{v}_{s}^{0}=\mathbf{1}, v^{0}=0$.}
            \STATE {\!\!\!\!\!\!\textbf{Step 1: Long-timescale optimization at the end of frame $t\!\in\![1, T]$}. }
            \begin{itemize}[leftmargin=0.35cm, itemindent=0cm]
              \item Acquire a statistical high-dimensional channel sample $\mathbf{H}^{t}\in \mathbb{C}^{N_{\mathrm{t}}\times K}$.
              \item Update the surrogate function according to \eqref{surrogate} based on $\mathbf{H}^{t}, \bm{\theta}^{t}, \mathbf{s}^{t}$.
              \item Calculate $(\bm{\bar{\theta}}^{t}, \mathbf{\bar{s}}^{t})$ iteratively in \eqref{eq:obj_final} based on Algorithm 2.
              \item Update $(\bm{\theta}^{t+1}, \mathbf{s}^{t+1})$ according to \eqref{update_theta} and \eqref{update_s}.
              \item Construct analog beamformer $\mathbf{F}_{\mathrm{RF}}^{t+1}=\mathbf{A}^{t+1}\mathbf{S}^{t+1}$.
              \item Let $t:=t+1$ and return to \textbf{Step 1}.
            \end{itemize}
             \STATE {\!\!\!\!\!\!\textbf{Step 2: Short-timescale optimization at the start of time-slot $t_{\mathrm{s}}\in[(t-1)T_{\mathrm{s}}+1, tT_{\mathrm{s}}]$}.}
                \begin{itemize}[leftmargin=0.35cm, itemindent=0cm]
                    \item Observe an instantaneous low-dimensional effective channel sample $\mathbf{H}_{\mathrm{e}}^{t_{\mathrm{s}}}\in \mathbb{C}^{K\times N_{\mathrm{RF}}}$.
                    \item With fixed analog beamformer $\mathbf{F}_{\mathrm{RF}}^{t}$, calculate the digital beamformer matrix $\mathbf{F}_{\mathrm{BB}}^{t_{\mathrm{s}}}$ according to \eqref{digital beamformer 2} and
                    \eqref{eq:digital beamformer 2_power}.
                    \item Let $t_{\mathrm{s}}:=t_{\mathrm{s}}+1$ and return to \textbf{Step 2}.
                \end{itemize}
    \end{small}
    \end{algorithmic}
\end{algorithm}

It can be observed that the newly constructed objective function \eqref{eq:obj_Boolean_b} contains two concave quadratic terms related to the variable $\mathbf{s}$. Hence, we attempt to transform the two concave terms into convex and solvable forms. To be more specific, we employ the MM approach to convert the quadratic function into a linear form, and then obtain the solution through the iteration \cite{MM}, \cite{R. Liu JSTSP}. The linear surrogate function for the first quadratic penalty term $\mathbf{s}^{T}(\mathbf{1}-\mathbf{s})$ at the current point $\mathbf{s}^{(m)}$ in the $m$-th iteration can be expressed as
\begin{equation}
  \label{eq:Boolean_MM_a}\mathbf{s}^{T}(\mathbf{1}-\mathbf{s})\leq -2(\mathbf{s}^{(m)})^{T}\mathbf{s}+(\mathbf{s}^{(m)})^{T}\mathbf{s}^{(m)} + \mathbf{s}^{T}\mathbf{1},
\end{equation}
which is a convex term and can be easily tackled. Similarly, the linear surrogate function for the second quadratic penalty term $\mathbf{\tilde{s}}^{T}(\mathbf{1}-\mathbf{\tilde{s}})$ can be given by
\begin{equation}
  \label{eq:Boolean_MM_b}\mathbf{\tilde{s}}^{T}(\mathbf{1}-\mathbf{\tilde{s}})\leq -2(\mathbf{\tilde{s}}^{(m)})^{T}\mathbf{\tilde{s}}+(\mathbf{\tilde{s}}^{(m)})^{T}\mathbf{\tilde{s}}^{(m)} + \mathbf{\tilde{s}}^{T}\mathbf{1},
\end{equation}
which exhibits discontinuity due to the $l_{0}$-norm involving in the inner function of $\mathbf{\tilde{s}}$.

To tackle the discontinuity issue of \eqref{eq:Boolean_MM_b}, we employ a logarithmic approximation function to address the discreteness of the $l_{0}$-norm. Specifically, we can approximate the $l_{0}$-norm, i.e., $\|x\|_{0}$, with a continuous and smooth function as
\begin{equation}\label{eq:log}
g_{\epsilon}(x)=\log\left(1+{x}/{\epsilon}\right)/\log\left(1+{1}/{\epsilon}\right),
\end{equation}
where $\epsilon>0$ is a tuning parameter that influences the degree of smoothness. A smaller $\epsilon$ has a lower approximation error at the expense of worse smoothness of the approximation function. According to \eqref{eq:log}, the $n$-th element of $\mathbf{\tilde{s}}$ (i.e., $\tilde{s}_{n}$) can be approximated as
\begin{equation}\label{eq:log_s}
\tilde{s}_{n} \approx\sum_{l=1}^{N_{\mathrm{RF}}}\log\left(1+{s_{n,l}}/{\epsilon}\right)/\log\left(1+{1}/{\epsilon}\right), ~\forall n,s_{n,l}\in[0,1].\!\!
\end{equation}

However, the logarithmic function \eqref{eq:log_s} is still concave with respect to $s_{n,l}$. Thus, we investigate to approximate it as a linear function at the current iterative point $s_{n,l}^{(m)}$ in \eqref{eq:log_s_linear} as shown at the bottom of this page. As a result, $\mathbf{\tilde{s}}$ is the linear function of $\mathbf{s}$, which can be written in a more concise representation shown in \eqref{eq:log_s_linear_matrix} also at the bottom of this page, where for brevity we define
\setcounter{equation}{49}
\begin{subequations}\begin{align}
 \mathbf{T} &\triangleq \mathbf{1}_{ N_{\mathrm{RF}}\times 1}^{T}\otimes\mathbf{I}_{N_{\mathrm{t}}},\\
  \mathbf{u}^{(m)} &\triangleq \frac{1}{\log(1+{1}/{\epsilon})}\log\big(1+{\mathbf{s}^{(m)}}/{\epsilon}\big), \\
  \mathbf{U}^{(m)} &\triangleq \frac{1}{\log(1+{1}/{\epsilon})}\mathrm{diag}^{-1}\big\{\epsilon+\mathbf{s}^{(m)}\big\}.
\end{align}\end{subequations}
After substituting the $l_{0}$-norm by a logarithmic function, and then locally approximating it by a linear function, the right hand of the inequality \eqref{eq:Boolean_MM_b} can be successfully converted into a continuous and convex form as
\begin{equation}
    \big(\mathbf{1}\!-\!2\mathbf{\tilde{s}}^{(m)}\big)^{T}\mathbf{T}\big(\mathbf{u}^{(m)}\!\!+\!\mathbf{U}^{(m)}(\mathbf{s}\!-\!\mathbf{s}^{(m)})\big)+(\mathbf{\tilde{s}}^{(m)})^{T}\mathbf{\tilde{s}}^{(m)}.
\end{equation}

In summary, based on the above derivations, in the $t$-th recursion, the analog beamformer design problem \eqref{eq:obj_Boolean_a} can be resolved by iteratively solving the following optimization problem
\begin{subequations}
\label{eq:obj_final}
\begin{align}
    \label{eq:obj_final_a}
    \begin{split}
    \!\!\!\!\!\!\!\!\!&\min_{\bm{\theta},\mathbf{s}}\ f^{t}(\bm{\theta},\mathbf{s}) +\varrho_{1}\big((\mathbf{1}\!-\!2\mathbf{s}^{(m)})^{T}\mathbf{s}+(\mathbf{s}^{(m)})^{T}\mathbf{s}^{(m)}\big)\\
    &+\!\varrho_{2}\Big((\mathbf{1}\!-\!2\mathbf{\tilde{s}}^{(m)})^{T}\mathbf{T}\big(\mathbf{u}^{(m)}\!\!+\!\!\mathbf{U}^{(m)}(\mathbf{s}\!-\!\mathbf{s}^{(m)})\big)\!\!+\!\!(\mathbf{\tilde{s}}^{(m)})^{T}\mathbf{\tilde{s}}^{(m)}\Big)
    \end{split}
    \\ &\ \mathrm{s.t. }\label{eq:obj_final_b}\ \ s_{n,l}\in[0,1],~ \forall n,l,
\end{align}
\end{subequations}
which is a convex problem and can be solved by many existing methods or convex optimization solvers, such as CVX. The inner iterative algorithm for solving $\bm{\bar{\theta}}^{t}$ and $\mathbf{\bar{s}}^{t}$ is summarized in Algorithm 2.

Finally, after acquiring $\bm{\bar{\theta}}^{t}$ and $\mathbf{\bar{s}}^{t}$ as the inner iterative algorithm converges, $\bm{\theta}$ and $\mathbf{s}$ can be respectively updated for the next frame as
\begin{subequations}\begin{align}\label{update_theta}
   \bm{\theta}^{t+1} &= (1-\rho^{t})\bm{\theta}^{t} + \rho^{t}\bm{\bar{\theta}}^{t},\\
\label{update_s}
   \mathbf{s}^{t+1} &= (1-\rho^{t})\mathbf{s}^{t} + \rho^{t}\mathbf{\bar{s}}^{t}.
\end{align}\end{subequations}
Furthermore, with the obtained $\bm{\theta}^{t+1}$ and $\mathbf{s}^{t+1}$, the analog beamformer can be constructed as $\mathbf{F}_{\mathrm{RF}}^{t+1}=\mathbf{A}^{t+1}\mathbf{S}^{t+1}$ for the next frame.

\subsection{Short-timescale Digital Beamformer Design}
In the $t_{\mathrm{s}}$-th short time-slot of the $t$-th frame, with the fixed analog beamformer $\mathbf{F}_{\mathrm{RF}}^{t}$, the digital beamformer is updated relying on the low-dimensional effective CSI $\mathbf{H}_{\mathrm{e}}^{t_{\mathrm{s}}}=(\mathbf{H}^{t_{\mathrm{s}}})^{H}\mathbf{F}_{\mathrm{RF}}^{t}\in \mathbb{C}^{K\times N_{\mathrm{RF}}}$, which can be obtained by processing pilot signals in the digital domain. Particularly, we adopt the classic minimum mean square error (MMSE) digital beamformer as
\begin{equation}\label{digital beamformer 2}
   \mathbf{\widetilde{F}}_{\mathrm{BB}}^{t_{\mathrm{s}}} =(\mathbf{H}_{\mathrm{e}}^{t_{\mathrm{s}}})^{H}\left( \mathbf{H}_{\mathrm{e}}^{t_{\mathrm{s}}}(\mathbf{H}_{\mathrm{e}}^{t_{\mathrm{s}}})^{H} + \sigma_{k}^{2}\mathbf{I}_{K}\right)^{-1},
\end{equation}
which should be further processed by a simple normalization operation to satisfy the power constraint as
\begin{equation}
    \label{eq:digital beamformer 2_power}
    \mathbf{F}_{\mathrm{BB}}^{t_{\mathrm{s}}}=\frac{\sqrt{P_{\mathrm{t}}}\ \mathbf{\widetilde{F}}_{\mathrm{BB}}^{t_{\mathrm{s}}}}{\|\mathbf{F}_{\mathrm{RF}}^{t}\mathbf{\widetilde{F}}_{\mathrm{BB}}^{t_{\mathrm{s}}}\|_{F}}.
\end{equation}

\subsection{Summary, Complexity and Convergence Analysis}
Based on the above algorithm development, the two-timescale dynamic hybrid beamforming design is outlined in Algorithm 3. At the end of each frame $t$, the BS obtains a high-dimensional channel sample and updates the long-timescale analog beamforming matrix $\mathbf{F}_{\mathrm{RF}}$. Specifically, the phase-shift vector $\bm{\bar{\theta}}^{t}$ and the switch vector $\mathbf{\bar{s}}^{t}$ are iteratively updated until the objective value in \eqref{eq:obj_final_a} converges, and then $\mathbf{F}_{\mathrm{RF}}^{t+1}=\mathbf{A}^{t+1}\mathbf{S}^{t+1}$ for the next frame can be obtained based on $\bm{\theta}^{t+1}$ and $\mathbf{s}^{t+1}$ in \eqref{update_theta} and \eqref{update_s}.
During each frame, the low-dimensional effective CSI is frequently acquired at the start of each time-slot $t_{\mathrm{s}}$.
And then it is utilized to obtain the short-timescale MMSE digital beamforming matrix $\mathbf{F}_{\mathrm{BB}}$ by \eqref{digital beamformer 2} and \eqref{eq:digital beamformer 2_power}.

We will briefly analyze the computational complexity of the proposed two-timescale dynamic hybrid beamforming design. Regarding the SSCA algorithm, the computational complexity predominantly lies in calculating the surrogate function with the matrix multiplication operations. To be specific, in \eqref{surrogate}, the computational complexity required for $\mathbf{v}_{\theta}^{t}$, $\mathbf{v}_{s}^{t}$, and $ v^{t}$ are $\mathcal{O}\{N_{\mathrm{t}}^{3}K^{2}\}$, $\mathcal{O}\{N_{\mathrm{t}}^{3}K^{2}\}$,  and $\mathcal{O}\{N_{\mathrm{t}}^{2}K^{2}\}$, respectively. As a result, the total computational complexity of the stochastic algorithm is $\mathcal{O}(2N_{\mathrm{t}}^{3}K^{2}+N_{\mathrm{t}}^{2}K^{2})$. Concerning \eqref{eq:obj_final} which is an $N_{\mathrm{t}}N_{\mathrm{RF}}$-dimensional quadratic optimization problem with $N_{\mathrm{t}}N_{\mathrm{RF}}$ second-order cone (SOC) constraints, the total computational complexity of solving variables $\bar{\bm{\theta}}^{t}$ and $\bar{\mathbf{s}}^{t}$ is $\mathcal{O}\{\sqrt{N_{\mathrm{t}}N_{\mathrm{RF}}+1}N_{\mathrm{t}}N_{\mathrm{RF}}(N_{\mathrm{t}}^{2}N_{\mathrm{RF}}^{2}+N_{\mathrm{t}}N_{\mathrm{RF}}+1)\}$.

In terms of the convergence analysis, the convergence of the typical two-timescale algorithm has been proved in \cite{A. Liu two-stage}-\cite{Y. Cai 2020}. Owing to the space limitation of the paper, we omit the detailed proof of the convergence. Please refer to \cite{A. Liu two-stage}-\cite{Y. Cai 2020} for a comprehensive derivation.

\section{Simulation Results}
In this section, extensive simulation results are presented to demonstrate the effectiveness of the dynamic hybrid beamforming architecture and the proposed beamforming design algorithms to enhance the performance of the ELAA near-field communication system. Unless otherwise specified, the BS is equipped with $N_{\mathrm{t}} = 1500$ antennas and the center carrier frequency is set as $f_{\mathrm{c}}= 28$ GHz. Without loss of generality, we assume that the number of UEs $K$ is equal to the number of RF chains $N_{\mathrm{RF}}$. The $K=3$ UEs are randomly distributed in the near-field region. We use the polar coordinate to describe their locations. In particular, both the angles and distances of the UEs follow the uniform distribution in each frame, i.e., $\vartheta_{k}\sim {U}[\bar{\vartheta}_{k}-\Delta\vartheta/2, \bar{\vartheta}_{k}+\Delta\vartheta/2]$ with an angle spread $\Delta\vartheta={\pi}/{48}$, and $r_{k}\sim {U}[\bar{r}_{k}-\Delta r/2, \bar{r}_{k}+\Delta r/2]$ with a distance spread $\Delta r=1$m, respectively. In each super-frame, the center angle and distance are changing among $\bar{\vartheta}_{k}\in[-{\pi}/{3},{\pi}/{3}]$ and $\bar{r}_{k}\in[2\mathrm{m},5\mathrm{m}]$, respectively. The channel gain involves typical distance-dependent path loss and the radiating gain. The former one is modeled as $L_{k,n}=10^{-\frac{C_{0}}{10}}(\frac{r_{k,n}}{D_{0}})^{-\alpha}$ for the link between the $k$-th UE and the $n$-th antenna with $C_{0}=30$dB, $D_{0}=1$m, and $\alpha=3$. The latter one is formulated as $G_{k,n}=(\cos\vartheta_{k,n})^{3}$, with $\vartheta_{k,n}\in (-{\pi}/{2},{\pi}/{2})$. The noise power at UEs is set as $-80$dBm. Additionally, in the case of two-timescale dynamic hybrid beamforming design, we assume that there are $T=120$ frames in each super-frame, and each frame contains $T_{\mathrm{s}}=200$ time-slots.
In the 6G networks, the duration of a super-frame referred to as channel statistics coherence time is generally spanning from a few seconds to a few minutes, while the duration of a time-slot termed channel coherence time is in the range of microseconds to a few milliseconds \cite{superframe}.

\begin{figure}[t]
    \centering
    \vspace{-0.3cm}
    \subfigure[Real-time design.]{\includegraphics[width= 1.72 in]{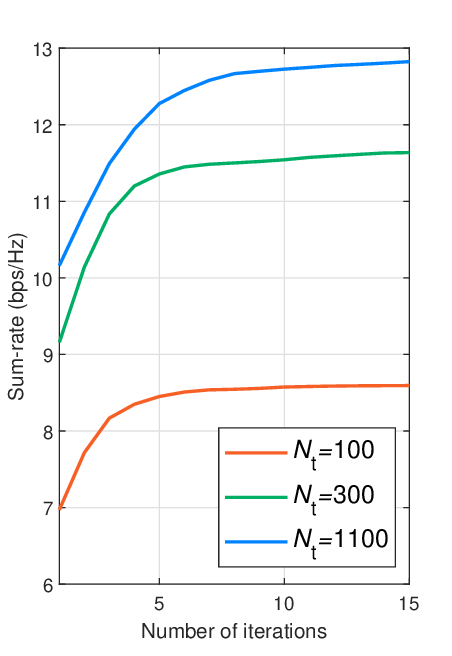}}%
    \vspace{-0.0 cm}
    \subfigure[Two-timescale design.]{\includegraphics[width= 1.72 in]{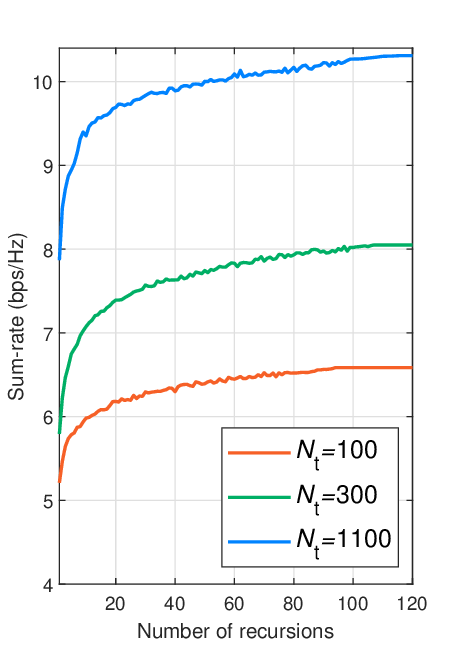}}
    \caption{Convergence performance of the proposed dynamic hybrid beamforming design algorithms.}
     \vspace{-0.3 cm}
    \label{fig:iteration}
\end{figure}

\begin{figure}[t]
   \hspace{-0.5 cm}
  \includegraphics[width= 3.8 in]{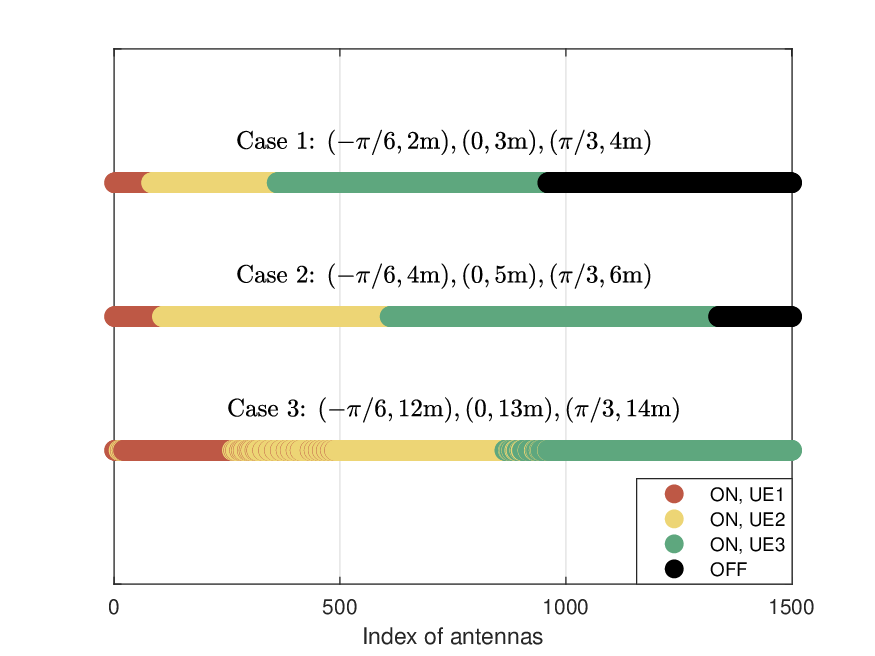}
  \caption{Illustration of antenna selection (three UEs under three cases with different locations).   \\}
  \label{selection}
  \vspace{-0.3 cm}
\end{figure}

\subsection{Convergence of Two Algorithms}
The convergence performance of the proposed dynamic hybrid beamforming designs is presented in Fig. \ref{fig:iteration}. It should be emphasized that the implications of convergence are distinct for the two algorithms. For the real-time case, convergence refers to the iterative calculation of the analog beamformer and digital beamformer in each time-slot given instantaneous CSI, while for the two-timescale case, it refers to the recursive learning of the channel statistics based on the high-dimensional channel samples and updating the analog beamformer in each frame. The results in Fig. \ref{fig:iteration} demonstrate that both the proposed algorithms can achieve satisfactory convergence speed, suggesting a favorable level of computational complexity. Moreover, we observe that the two-timescale design achieves satisfactory performance when $T=120$ frames are utilized for updating the analog beamformer in the long-timescale.

\subsection{Illustration of Antenna Selection}
In Fig. \ref{selection}, we illustrate the results of antenna selection under different scenarios. Particularly, we evaluate three cases in which the distances between the ELAA and three UEs gradually  increase.
Since the two-timescale algorithm is designed based on the statistical CSI of UEs, we only illustrate the antenna selection based on the real-time dynamic hybrid bemforming design, which can directly reflect the relationship between antenna selection and UE locations.
As shown in Fig. \ref{selection}, for Case 1 with the smallest distance between the ELAA and the UEs, the near-field effect is significant. Consequently, about $37\%$ ``ineffective" antenna elements with poor near-field channel gain are switched to the ``off mode'' for power saving and performance enhancement.
When the distance becomes relatively larger for Case 2,
more antennas should be activated.
For Case 3, where the distance is sufficiently large, the channel begins to exhibit far-field characteristics, and all antennas must be utilized to serve the UEs. Thus, the phenomenons in Fig. \ref{selection} illustrate the impact of near-field effect on the utilization of ELAA and verify the effectiveness of our proposed algorithm.

\begin{table}[t]
\caption{\textsc{Total Power Consumption for Different Architectures in} mm\textsc{Wave Systems.}}
\begin{center}
\begin{tabular}{cc}
   \toprule
    {  \small Architecture}&\qquad { \small Total Power Consumption $P_\mathrm{tot}$}\\
    \midrule
    \small Fully-digital &\small\tabincell{l}{$P_{\mathrm{t}}+P_\mathrm{BB}+N_\mathrm{t}P_\mathrm{RF}$} \\
    \vspace{1mm}
    \small Fully-connected &\small\tabincell{l}{$P_{\mathrm{t}}+P_\mathrm{BB}+N_\mathrm{RF}P_\mathrm{RF}+N_\mathrm{t}N_\mathrm{RF}P_\mathrm{PS}$} \\
    \vspace{1mm}
    \small Fixed-subarray &\small\tabincell{l}{$P_{\mathrm{t}}+P_\mathrm{BB}+N_\mathrm{RF}P_\mathrm{RF}+N_\mathrm{t}P_\mathrm{PS}$}\\
    \vspace{1mm}
    \small Dynamic-subarray &\small\tabincell{l}{$P_{\mathrm{t}}+P_\mathrm{BB}+N_\mathrm{RF}P_\mathrm{RF}+MP_\mathrm{PS}+MP_\mathrm{SW}$} \\
    \bottomrule
\hline
\end{tabular}
\label{table:1}
\end{center}
\end{table}

\subsection{Impact of Transmit Power}
\begin{figure}[t]
  \centering
  \includegraphics[width= 3.5 in]{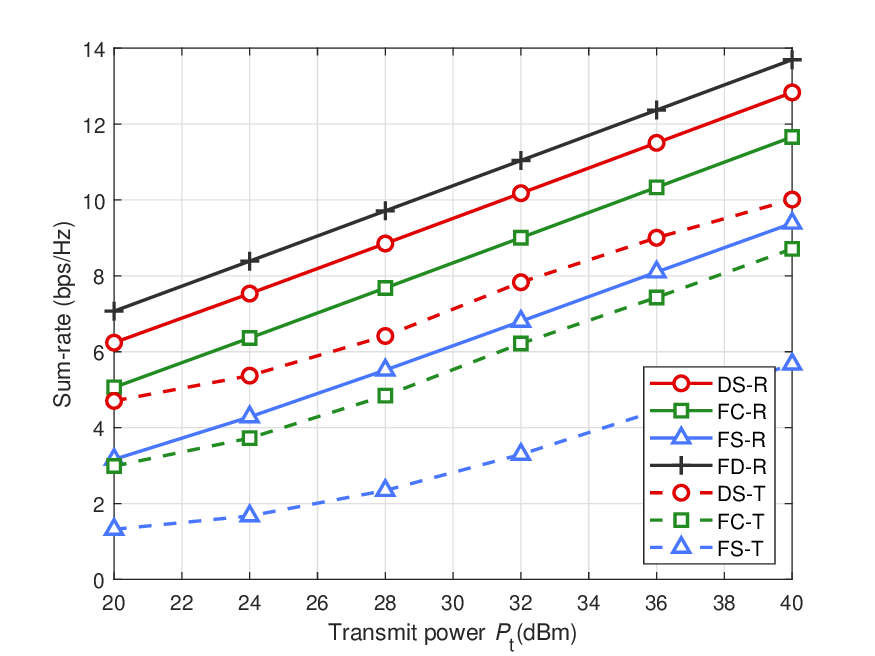}
  \caption{The sum-rate versus the transmit power $P_\text{t}$.}
  \label{fig:fig1}
   \vspace{0.43 cm}
\end{figure}

In Fig. \ref{fig:fig1}, we show the sum-rate performance versus transmit power under different beamforming architectures for real-time and two-timescale frameworks. Especially, the real-time fully-digital beamforming design denoted as ``FD-R" is included as a performance upper bound. The widely considered fully-connected and fixed-subarray hybrid beamforming architectures are also evaluated for comparison purposes, which are denoted as ``FC-R" and ``FS-R" under the real-time framework, and ``FC-T" and ``FC-T" under the two-timescale framework, respectively. The proposed real-time dynamic hybrid beamforming design with instantaneous CSI and the proposed two-timescale dynamic hybrid beamforming design with channel statistics are denoted as ``DS-R" and ``DS-T", respectively.
It can be observed from Fig. \ref{fig:fig1} that, both the proposed ``DS-R" and ``DS-T" algorithms can consistently provide better performance compared with their competitors. This result verifies that the dynamic hybrid beamforming can provide superior performance for ELAA near-field communications thanks to its advanced flexibility of ``on/off" mode switch and subarray selection in the analog beamforming design. In addition, we notice that the proposed two-timescale beamforming design can significantly reduce the overhead associated with channel estimation at the price of certain performance loss, which offers a satisfactory balance between performance and overhead for ELAA near-field communication systems.

\begin{figure}[t]
  \centering
  \includegraphics[width= 3.5 in]{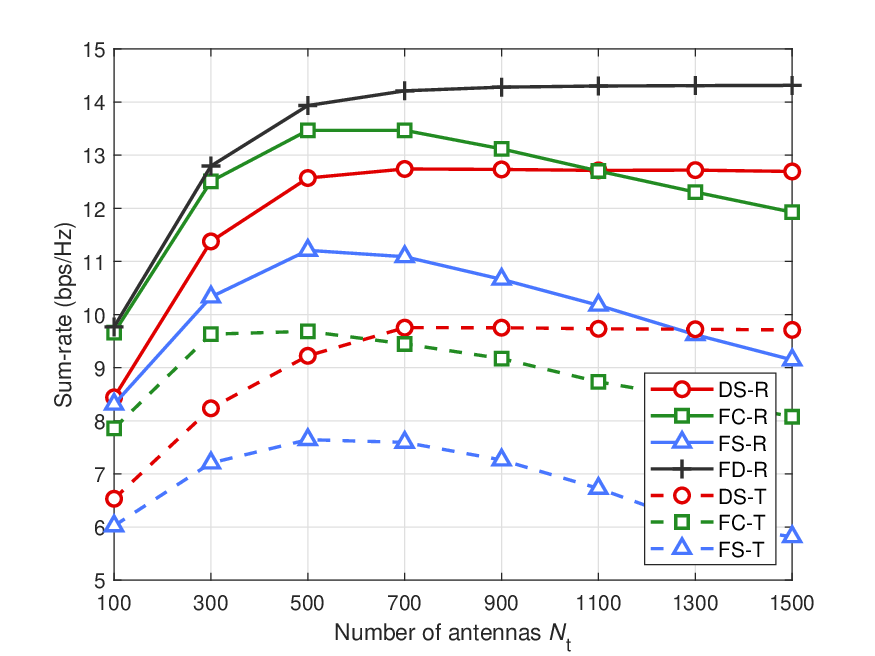}
  \caption{The sum-rate versus the number of transmit antennas $N_{\mathrm{t}}$ ($P_{\mathrm{t}}=40 \mathrm{dBm}$).}\label{fig:fig2}
   \vspace{0.2 cm}
\end{figure}

\begin{figure}[t]
  \centering
  \includegraphics[width= 3.5 in]{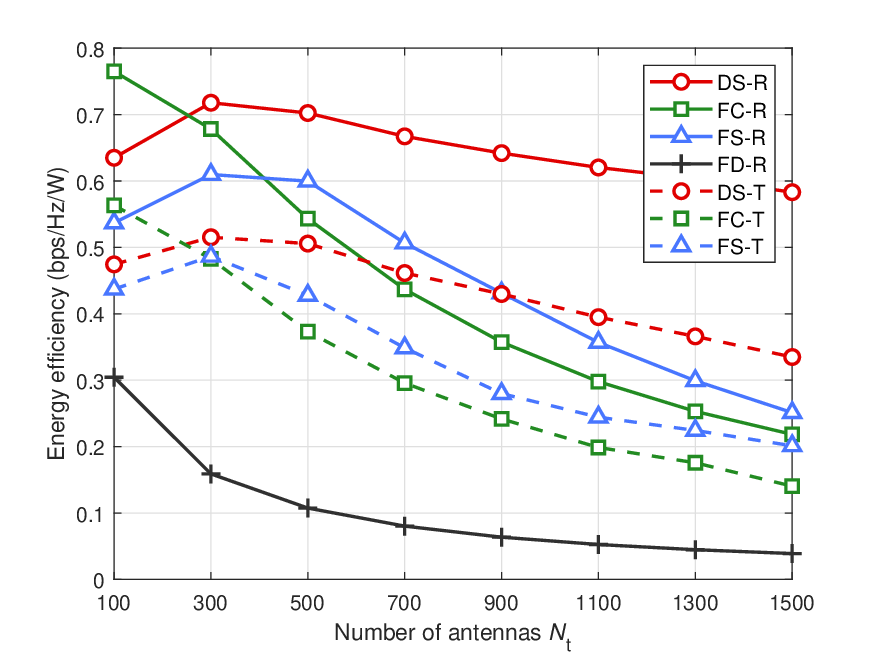}
  \caption{The energy efficiency versus the number of transmit antennas $N_{\mathrm{t}}$ ($P_{\mathrm{t}}=40 \mathrm{dBm}$).}\label{fig:fig3}
\end{figure}

\begin{figure}[t]
  \centering
  \includegraphics[width= 3.5 in]{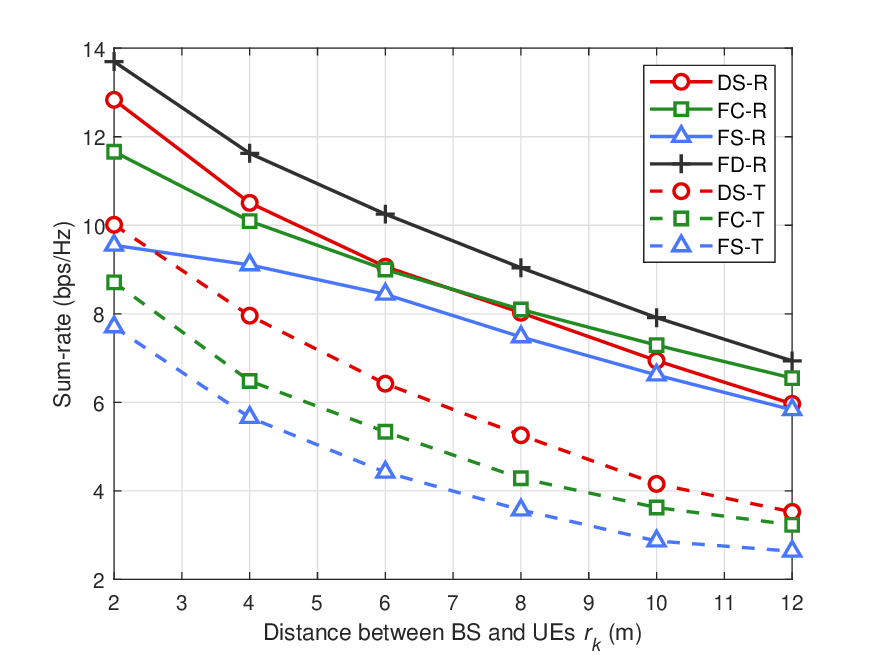}
  \caption{The sum-rate versus the distance between BS and UEs.}\label{fig:fig5}
  \vspace{0.4cm}
\end{figure}

\subsection{Impact of Number of Transmit Antennas}
Fig. \ref{fig:fig2} depicts the sum-rate of different beamforming architectures versus the number of transmit antennas. It is noted that, for the fully-connected and fixed-subarray schemes, more antennas may even lead to a decreased sum-rate in both real-time and two-timescale circumstances. This phenomenon indicates that simply equipping more antennas does not always provide sum-rate performance improvement in ELAA near-field communication systems if the non-uniform channel gains across different antennas are not properly handled. Our proposed dynamic hybrid beamforming structure effectively addresses this issue and guarantees a satisfactory sum-rate in real-time or two-timescale frameworks.

On the other hand, an ELAA system with more antennas requires more PSs and switches, leading to higher power consumption. Next, we further evaluate the energy efficiency defined as
\begin{equation}
\eta=\frac{\sum_{k=1}^{K}R_{k}}{P_{\mathrm{tot}}},
\end{equation}
where $P_{\mathrm{tot}}$ is the total power consumption. The total power consumption $P_{\mathrm{tot}}$ under different structures is summarized in Table \ref{table:1}, where $P_{\mathrm{t}}$ is the transmit power, $M$ represents the number of antennas used in the dynamic structure, and $P_\mathrm{BB}$, $P_\mathrm{RF}$, $P_\mathrm{PS}$, and $P_\mathrm{SW}$ are the powers consumed by the baseband processor, one RF chain, one PS, and one switch, respectively. In this paper, we use the practical values $P_\mathrm{BB}=200$mW, $P_\mathrm{RF}=250$mW, $P_\mathrm{PS}=10$mW, and $P_\mathrm{SW}=5$mW \cite{H. Li 2020 Dynamic}. Fig. \ref{fig:fig3} illustrates the energy efficiency versus the number of antennas. It can be observed that both the proposed ``DS-R" and ``DS-T" schemes exhibit superior energy efficiency in comparison with the fully-digital beamforming schemes and the other two hybrid beamforming competitors, especially for a larger $N_\text{t}$.

\subsection{Impact of UEs' Location}
The sum-rate versus the distance between BS and UEs is shown in Fig. \ref{fig:fig5}. It is obvious that as the distance increases, the performance gradually deteriorates due to the higher path loss. More importantly,  since the near-field impact diminishes as the distance increases, the performance improvement of dynamic hybrid beamforming designs becomes less significant, but
it  is consistently superior to that of the fixed-subarray scheme.
This result demonstrates the advantages of the proposed dynamic architecture, which provides substantial performance improvements for near-field scenarios while guaranteeing satisfactory performance for far-field cases.
Moreover, based on the above observations and conclusions, we can infer that the proposed dynamic hybrid beamforming designs can still exhibit great performance in a near-field far-field hybrid scenario.

\begin{figure}[t]
    \centering
    \vspace{-0.3cm}
    \subfigure[Angle spread with $\Delta r=0$.]{\includegraphics[width= 1.72 in]{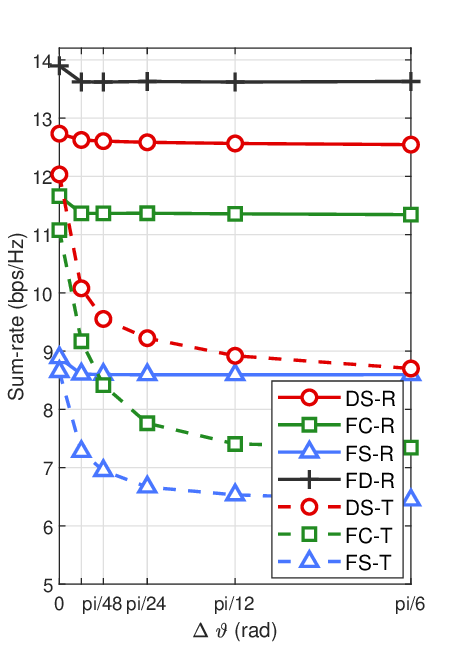}}%
    \subfigure[Distance spread with $\Delta\vartheta\!=\!0$.]{\includegraphics[width= 1.72 in]{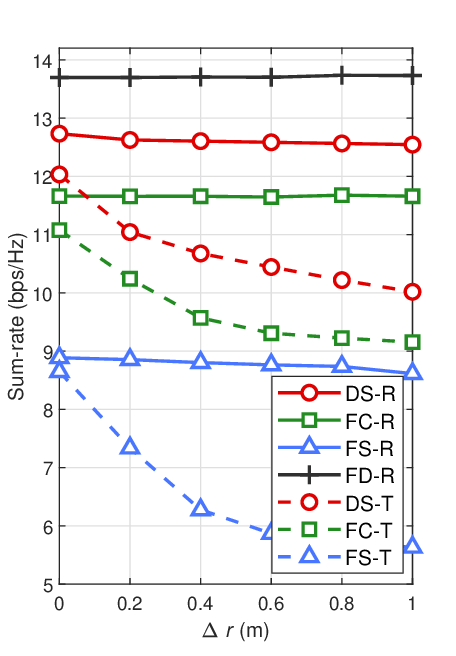}}
    \caption{The sum-rate versus the angle/distance spread.}
    \label{fig:delta}
\end{figure}

\begin{figure}[t]
  \centering
  \hspace{-0.6cm}
  \includegraphics[width= 3.5 in]{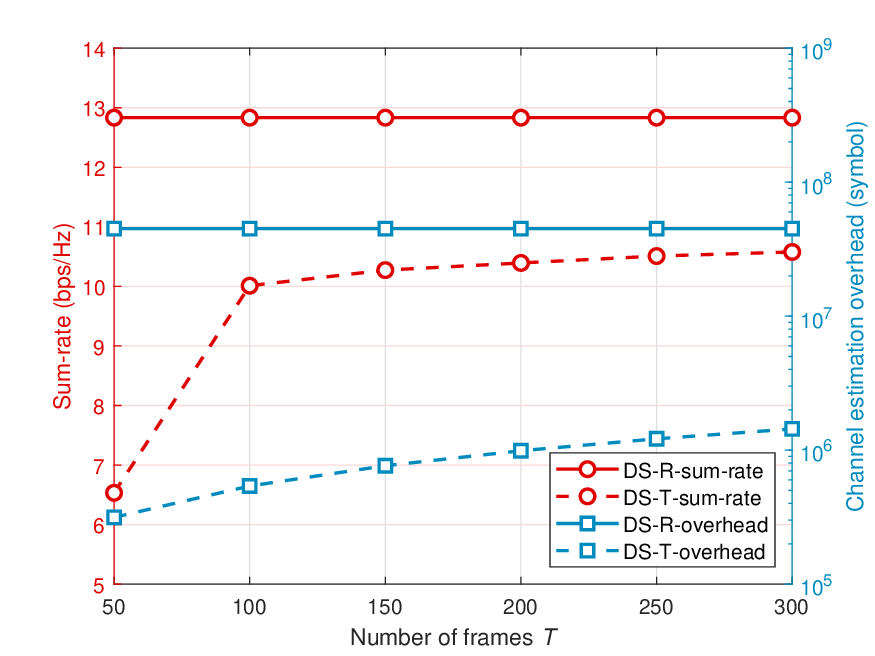}
  \caption{The sum-rate and channel estimation overhead versus the number of frames $T$.}
  \label{fig:T}
\end{figure}

\subsection{Impact of Angle and Distance Spread}
We note that the proposed two-timescale dynamic hybrid beamforming design algorithm is essentially a stochastic approach that is highly dependent on the channel statistics.
Thus, the mobility of UEs may cause the variation of channel statistics, which will potentially deteriorate the performance.
To measure this impact, in Fig. \ref{fig:delta} we illustrate the sum-rate versus the mobility of UEs in terms of angle and distance spread of their locations.
The performance for real-time cases is also included as a benchmark.
We observe that when UEs have a larger angle or distance spread (i.e., higher mobility), acquiring the channel statistics becomes more difficult, thereby causing an inevitable deterioration in performance. Besides, when both the angle and distance of the UEs remain unchanged, i.e., $\Delta\vartheta=\Delta r=0$, the performance of the two-timescale framework is very close to that of the real-time framework.

\subsection{Impact of Channel Sampling Rate}
For the two-timescale dynamic hybrid beamforming design, a higher high-dimensional channel sampling/estimation rate potentially provides better sum-rate performance owing to more accurate and rapid learning of channel statistics, but it will lead to a higher channel estimation overhead. To numerically illustrate the impact of the channel sampling rate on both performance and overhead, we fix the length of the super-frame and change the number of frames within each super-frame. A larger number of frames implies a higher rate of channel sampling.
Fig. \ref{fig:T} shows both the sum-rate performance and channel estimation overhead of the proposed algorithms versus the number of frames within each super-frame.
Specifically, we assume that the typical least-square channel estimation is employed for all algorithms. Thus the required channel estimation overhead is proportional to the number of channel coefficients, as analyzed in Table II. It can be observed from both Fig. \ref{fig:T} and Table II that the two-timescale scheme significantly reduces the channel estimation overhead compared to the real-time scheme.
More importantly, as the number of frames (i.e., channel sampling rate) increases, the performance of the two-timescale design improves and eventually maintains a satisfactory level, but with a gradual increase in channel estimation overhead.
Moreover, we notice a good trade-off between performance and overhead with the number of frames being $T=120$, which is thus used in previous simulation studies.

\begin{table}[t]
\caption{\textsc{Channel Estimation Overhead Comparison in One Super-frame}}
\begin{center}
\begin{tabular}{cccc}
   \toprule
    {  \small }&{ \small Analog }&{ \small Digital}& { \small Total}\\
    \midrule
    \small Real-time &\small\tabincell{l}{$N_{\mathrm{t}}KTT_{\mathrm{s}}$} &\small\tabincell{l}{0}&\small\tabincell{l}{$N_{\mathrm{t}}KTT_{\mathrm{s}}$}\\
    \vspace{1mm}
    \small Two-timescale &\small\tabincell{l}{$N_{\mathrm{t}}KT$} &\small\tabincell{l}{$K\!N_{\mathrm{RF}}TT_{\mathrm{s}}$}&\small\tabincell{l}{$N_{\mathrm{t}}KT\!+\!K\!N_{\mathrm{RF}}TT_{\mathrm{s}}$}\\
    \bottomrule
\hline
\end{tabular}
\vspace{-0.0cm}
\label{table:2}
\end{center}
\end{table}

\section{Conclusions}
In this paper, we proposed a novel dynamic hybrid beamforming architecture to alleviate the near-field effect and boost performance for ELAA systems, in which each antenna is adaptively selected to form non-overlapping subarrays for transmission or not to be used. When the instantaneous CSI can be acquired during each channel coherence time, a real-time dynamic hybrid beamforming design was proposed to maximize the achievable sum-rate under the constraints of PSs, dynamic connection network, and total power. An efficient FP-manifold-based algorithm was developed to solve the non-convex optimization problem.
When the instantaneous CSI cannot be easily obtained in real-time, a two-timescale dynamic hybrid beamforming design was proposed to maximize the ergodic sum-rate under the same constraints. We developed an SSCA-MM-based algorithm to solve this stochastic non-convex problem.
The simulation findings indicated that the dynamic hybrid beamforming architecture offers superior performance under both real-time and two-timescale frameworks, making it a more favorable choice for ELAA systems.
Considering the significant computational complexity associated with the large number of antennas in ELAA systems, our future research will concentrate on decentralized beamforming design approaches to achieve computational efficiency. In this distributed architecture, the antennas will be partitioned into multiple clusters to facilitate low-complexity computation, while central data fusion will be employed to ensure overall system consensus.

\end{document}